\documentclass[prd,twocolumn,superscriptaddress,nofootinbib]{revtex4-1}
\usepackage{epsfig}
\usepackage{graphicx}
\usepackage[english]{babel}
\usepackage{hyphenat}
\usepackage{amsmath}
\usepackage{amssymb}
\usepackage{mathtools}
\usepackage{mathrsfs}
\usepackage{slashed}
\usepackage{epstopdf}
\usepackage[dvipsnames]{xcolor}
\usepackage{booktabs}
\usepackage{subfigure}
\usepackage{float}
\usepackage{url}
\usepackage{breakurl}

\definecolor{lcolor}{rgb}{0.,0.0,0.}
\definecolor{citcolor}{rgb}{0,0.,0.5}
\usepackage[breaklinks,colorlinks,urlcolor=blue,citecolor=blue,linkcolor=blue]{hyperref}
\usepackage{multirow}
\usepackage{ltablex}
\usepackage{soul}
\usepackage{braket}
\usepackage{hepunits}
\colorlet{darkgreen}{green!50!black}
\colorlet{darkblue}{blue!70!black}
\colorlet{brightyellow}{yellow!75!red}
\colorlet{orange}{red!50!yellow}
\colorlet{darkgray}{gray!50!black}

\newcommand*\diff{\mathop{}\!\mathrm{d}}
 \newcommand{\GeV}{{{\,}\textrm{GeV}}}
\def\rrangle{\rangle\!\rangle}
\def\llangle{\langle\!\langle}
\newcommand{\beq}{\begin{eqnarray}}
\newcommand{\eeq}{\end{eqnarray}}

\newcommand{\bem}{\begin{multline}}
\newcommand{\eem}{\end{multline}}
\newcommand{\beg}{\begin{gather}}
\newcommand{\eeg}{\end{gather}}

\newcommand{\ben}{\begin{eqnarray*}}
\newcommand{\een}{\end{eqnarray*}}

\newcommand{\eqn}[1]{Eq.~\eqref{#1}}

\def\cT{{\cal T}}

\newcommand{\secn}[1]{Section~1}
\newcommand{\appn}[1]{Appendix~1}

\long\def\comment#1{ }

\def\Tr{\text{Tr}}

\def\and{\quad\text{and}\quad}

\def\0{{\boldsymbol 0}}
\def\p{{\boldsymbol p}}

\def\k{{\boldsymbol k}}

\def\n{{\boldsymbol n}}
\def\x{{\boldsymbol x}}
\def\y{{\boldsymbol y}}

\newcommand{\Kceil}{\lceil K \rceil }

\begin{document}

\title{
Quantum simulation of in-medium QCD jets: momentum broadening, gluon production, and entropy growth 
}

\author{Jo\~ao Barata}
\email[]{jlourenco@bnl.gov}
\affiliation{Physics Department, Brookhaven National Laboratory, Upton, NY 11973, USA}
\author{Xiaojian Du}
\email[]{xiaojian.du@usc.es}
\affiliation{Instituto Galego de Fisica de Altas Enerxias (IGFAE), Universidade de Santiago de Compostela, E-15782 Galicia, Spain}
 \author{Meijian Li}
\email[]{meijian.li@usc.es}
\affiliation{Instituto Galego de Fisica de Altas Enerxias (IGFAE), Universidade de Santiago de Compostela, E-15782 Galicia, Spain}
\author{Wenyang Qian}
\email[]{qian.wenyang@usc.es}
\affiliation{Instituto Galego de Fisica de Altas Enerxias (IGFAE), Universidade de Santiago de Compostela, E-15782 Galicia, Spain}
\author{Carlos A. Salgado}
\email[]{carlos.salgado@usc.es}
\affiliation{Instituto Galego de Fisica de Altas Enerxias (IGFAE), Universidade de Santiago de Compostela, E-15782 Galicia, Spain}

\begin{abstract} 
Jets provide one of the primary probes of the quark-gluon plasma produced in ultrarelativistic heavy ion collisions and the cold nuclear matter explored in deep inelastic scattering experiments.
However, despite important developments in the last years, a description of the real-time evolution of QCD jets inside a medium is still far from being complete. In our previous work, we have explored quantum technologies as a promising alternative theoretical laboratory to simulate jet evolution in QCD matter, to overcome inherent technical difficulties in present calculations.
Here, we extend our previous investigation from the single particle $\ket{q}$ to the $\ket{q}+\ket{qg}$ Fock space, taking into account gluon production. 
Based on the light-front Hamiltonian formalism, we construct a digital quantum circuit that tracks the evolution of a multi-particle jet probe in the presence of a medium described as a stochastic color field.
Studying the momentum broadening of the jet state, we observe sizable sub-eikonal effects by comparing to eikonal estimates.
 We also study the medium-induced modifications to the gluon emission probability, which exhibit small corrections compared to the vacuum splitting function. 
In addition, we study the time evolution of the von-Neumann entropy associated with the quark component; 
we find that the exponential of the entropy grows linearly in time for the bare quark but super-linearly when taking into account gluon emission. 
\end{abstract}

\maketitle

%%%%%%%%%%%%%%%%%%
\section{Introduction}
\label{sec:intro}
%%%%%%%%%%%%%%%%%%

Heavy-ion collisions at the Relativistic Heavy-Ion Collider and the Large Hadron Collider, produce collimated particle showers originated from highly energetic quarks and gluons, known as jets, that evolve simultaneously with the hot and dense quark gluon plasma. A similar scenario is expected in future deep inelastic scattering experiments, where these sprays of particles traverse the cold nuclear matter target. 

Jets can resolve the underlying medium at different energy scales and thus offer an optimal probe to study the structure of QCD matter. Their evolution in these environments is characterized by sizeable modifications to the jets' structure, which are reflected at the level of the final state distributions. Phenomenologically, these result in the broadening of the transverse momentum distributions, due to instantaneous interactions with the medium, and the alteration of the radiation pattern, leading to an excess of energy flowing at large angles; for recent reviews on medium-induced jet modifications see~\cite{Casalderrey-Solana:2007knd, Majumder:2010qh, Qin:2015srf, Apolinario:2022vzg,Blaizot:2015lma}. To the present day, the theoretical study of all these effects has been mainly constrained to lower orders in perturbation theory~\cite{Gyulassy:2000er,Zakharov:1996fv,Baier:1996sk}, with a limited number of higher order calculations being available~\cite{Arnold:2015qya,Fickinger:2013xwa}. Compared to their vacuum counterpart, the slower progress seen in this research field is mainly tied to the highly complex multi-particle interference pattern determining parton fragmentation in matter. 

More recently, it has been argued that novel advances in quantum information science could be used to leverage our understanding of in medium jet physics~\cite{Barata:2021yri,DeJong:2020riy}.
In particular, future large-scale and fault-tolerant digital quantum computers can potentially offer a platform to efficiently simulate large quantum systems in real time, using the quantum simulation algorithm. These devices capture the dynamics of the target quantum system by evolving a controlled finite-dimensional quantum system whose dynamics can be engineered appropriately. Nonetheless, their practical implementation is still complex  and problem-dependent.

In our preceding work~\cite{Barata:2022wim}, we provided a quantum simulation protocol to simulate the real-time evolution of a single hard parton in the presence of a stochastic background. This can be thought of as describing the propagation of the leading parton inside the jet, and it allowed us to establish the basic aspects of this approach.
In this work, we extend the strategy to include gluon radiation, thus allowing the jet to have a non-trivial structure.

Our quantum formulation is based on a nonperturbative light-front Hamiltonian approach, the time-dependent basis light-front quantization (tBLFQ)~\cite{Zhao:2013cma}; see its various applications in Refs.~\cite{ Hu:2019hjx,Chen:2017uuq,Lei:2022nsk, Li:2020uhl, Li:2021zaw, Li:2023jeh}. 
This approach allows us to quantize the QCD Hamiltonian and perform real-time simulations at the amplitude level, which is natural and well-suited for applications in quantum computers.
In particular, we can exactly track the jet state in time and extract observables by computing the expectation values for appropriate operators. 

Quark jet evolution in a colored medium using tBLFQ has already been simulated on classical computers: first in the $\ket{q}$ Fock space~\cite{Li:2020uhl}, and later extended to the $\ket{q}+\ket{qg}$ space~\cite{Li:2021zaw, Li:2023jeh}.
These studies showed the interplay between coherence and multiple scattering in gluon emission from nonperturbative perspectives.
Our preceding work~\cite{Barata:2022wim} provides a quantum implementation in the $\ket{q}$ space and further investigations in jet momentum broadening. 
This work takes a step forward by extending to the $\ket{q}+\ket{qg}$ space.
We build a digital quantum circuit that can track the evolution of a jet state in the presence of a medium background field.
We focus on the momentum broadening of the jet state and its branching pattern, both in vacuum and medium. 
We also discuss modifications to the single particle entropy growth due to radiation. 

This manuscript is organized as follows. In Sec.~\ref{sec:theory}, we review the formulation of a jet evolution in a dense medium within the quark and quark-gluon Fock sectors using light-front Hamiltonian formalism; we then build the quantum simulation algorithm of the jet evolution. In Sec.~\ref{sec:results}, we present numerical results of our approaches via quantum simulation of the jet using {\tt Qiskit}. In Sec.~\ref{sec:conclusion}, we summarize our current results and discuss the future avenue of this work.

%%%%%%%%%%%%%%%%%%
\section{Methodology}
\label{sec:theory}
%%%%%%%%%%%%%%%%%%

In our preceding work, Ref.~\cite{Barata:2022wim}, we quantum simulated the evolution of a single particle through a colored medium. 
Our method was based on the tBLFQ, a numerical non-perturbative light-front Hamiltonian approach developed to study real-time problems; see Refs.~\cite{Li:2020uhl,Li:2021zaw} for further details and applications using classical methods.\footnote{See also Ref.~\cite{Barata:2021yri} for related discussions.} 
In this work, we extend the Fock space to $\ket{q}+\ket{qg}$, thus including gluon emission and absorption. 
On a classical computer, the simulation of this process has been studied in Refs.~\cite{Li:2021zaw, Li:2023jeh}, also within tBLFQ. 

In the following, we first briefly review the basics of tBLFQ to simulate the in-medium jet evolution process in the $\ket{q}+\ket{qg}$ Fock space in Sec.~\ref{sec:theory_c} and then detail how to apply this method using the quantum simulation algorithm
in Sec.~\ref{sec:to_qcomputer}.

\subsection{Jet evolution in the light-front Hamiltonian formalism of tBLFQ}\label{sec:theory_c}

\begin{figure}[thp!]
    \centering
    \includegraphics[width=0.35\textwidth]{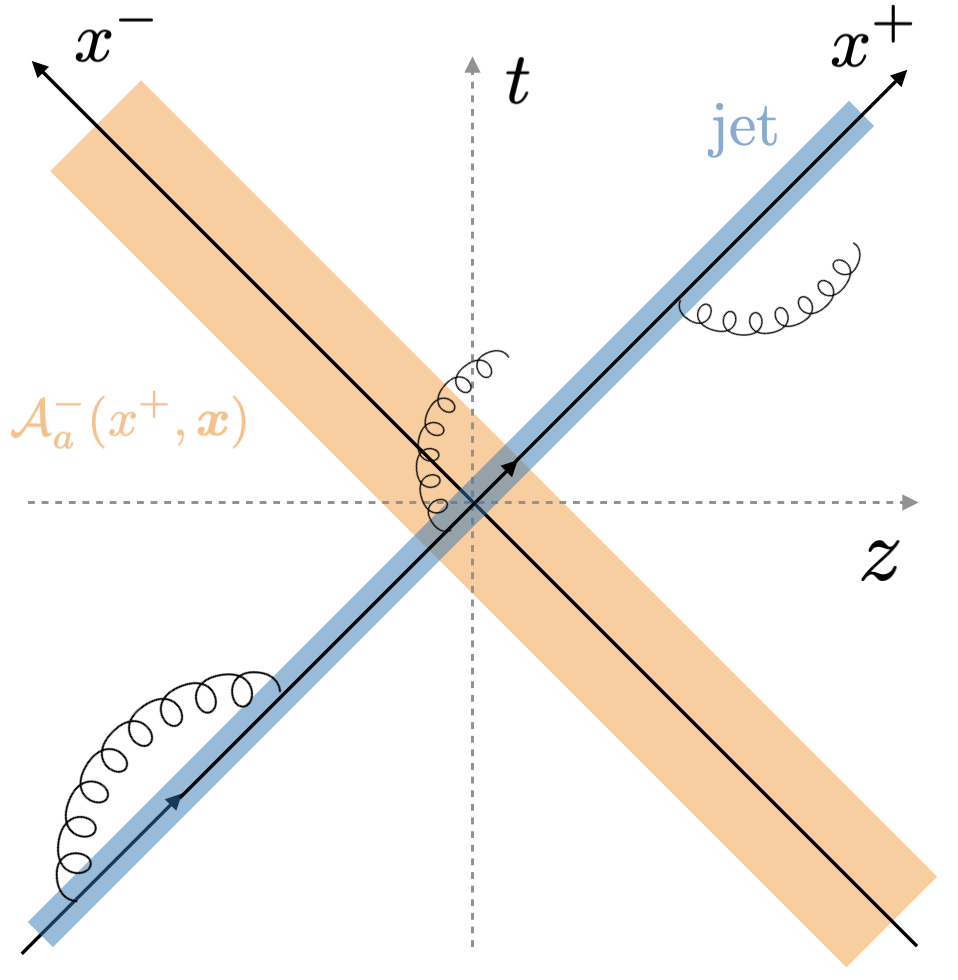}
    \caption{
    An illustration of the jet (blue line, dressed by helical lines representing the gluon in the $\ket{qg}$ state) evolution in the presence of a highly boosted background medium (orange band) described by a classical field $\mathcal{A}^\mu(x)$.
    }
    \label{fig:event_picture}
\end{figure}

We consider the propagation of a highly energetic massless jet with light-front momentum $p=(p^+, p^-, {\boldsymbol p})$, moving close to the light cone along the $x^+$ direction.\footnote{The light-front coordinates are defined as \( (x^+, \x, x^-) \), where \(x^+=x^0+ x^3\) is the light-front time,  \(x^-=x^0-x^3\) the longitudinal coordinate, and  \(\x=(x^1, x^2)\) the transverse coordinates. The letters in bold, such as $\x$, denote  transverse vectors, while their magnitude is denoted by $x_\perp\equiv |\x|$. 
The non-vanishing elements of the metric tensors $g^{\mu\nu}$ and $g_{\mu\nu}$ are, $ g^{+-}=g^{-+}=2$, $g_{+-}=g_{-+}=1/2$, $g^{ii}=g_{ii}=-1$ with $i=1,2$.} %(see App.~\ref{appendix:conventions} for conventions of coordinates in this paper).
This hard probe evolves in the presence of a dense medium, which can be boosted to move along the $x^-$ direction. The quark interacts with the medium over a finite distance in light-front time $x^+=[0,L_\eta]$. This process is illustrated in Fig.~\ref{fig:event_picture}. The dynamics of this system are set by the QCD Lagrangian in the presence of an external field,
\begin{align}\label{eq:Lagrangian}
 \mathcal{L}=-\frac{1}{4}{F^{\mu\nu}}_a F^a_{\mu\nu}+\overline{\Psi}(i\gamma^\mu  D_\mu -  m_q)\Psi\;,
\end{align}
where $F^{\mu\nu}_a\equiv\partial^\mu C^\nu_a-\partial^\nu C^\mu_a-g f^{abc}C^\mu_b C^\nu_c$ is the field strength tensor, $D^\mu\equiv \partial_\mu +ig C^\mu$ the covariant derivative, and $ C^\mu= A^\mu + \mathcal{A}^\mu$ is the sum of the quantum gauge field $ A^\mu$ and the background gluon field $\mathcal{A}^\mu$. 

In this work, we truncate the Fock space of the jet to the leading  two sectors, $\ket{q}$ and $\ket{qg}$, such that the full quantum state can be written as
\begin{align}\label{eq:Fock}
    \ket{\psi} = \psi_{q}\ket{q} + \psi_{qg} \ket{qg}\;,
\end{align}
where $\psi_{q}$ and $\psi_{qg}$ represent the respective Fock amplitudes. 
The light-front Hamiltonian can be obtained following the canonical light-front quantization formalism~\cite{Brodsky:1997de,Blaizot:2015lma,Li:2020uhl,Li:2021zaw} via the standard Legendre transformation, in the light-cone gauge of $A^+=\mathcal{A}^+=0$,
% and within the truncated Fock space, the light-front Hamiltonian consists of three parts, 
\begin{align}\label{eq:Hamiltonian_full}
P^-(x^+)=P_{KE}^- + V_\mathrm{qg}+ V_\mathcal{A}(x^+)\;.
\end{align}
Here, $P_{KE}^-$ stands for the kinetic energy part
\begin{align}
\begin{split}
       P_{KE}^-= &P_{KE,g}^- + P_{KE,q}^- \label{eq:Hamiltonian_kinetic}\\
    =&\int\diff x^-\diff^2 \x \bigg (
    -\frac{1}{2}A_a^j (i\nabla)^2_\perp A_j^a\\
    &
    +\frac{1}{2}\overline{\Psi}\gamma^+\frac{m^2-\nabla_\perp^2}{i\partial^+}\Psi
    \bigg )\;,
\end{split}
\end{align}
where $P_{KE,g}^-$ and $P_{KE,q}^-$ are the respective kinetic energies for the dynamical gluon and quark. 
The second term $V_{qg}$ is the interaction between the quark and gluon, %\xd{missing a coupling strength?}
\begin{align}
    V_{qg} =\int\diff x^-\diff^2 \x\,
    g \overline{\Psi}\gamma^\mu T^a\Psi A_\mu^a\;. \label{eq:Hamiltonian_Vqg}
\end{align}
The third term $V_\mathcal{A}(x^+)$ includes the
interaction of the background field with the quark and that
with the dynamical gluon,
\begin{align}\label{eq:Hamiltonian_VA}
\begin{split}
V_{\mathcal{A}}(x^+) =& V_{\mathcal{A}, q}(x^+) +  V_{\mathcal{A}, g}(x^+)\\
=&\int\diff x^-\diff^2 \x
\bigg ( g \bar\Psi\gamma^+ T^a\Psi \mathcal{A}_+^a (x^+)\\
&
+ g f^{abc}\partial^+A^c_i A^{bi} \mathcal{A}^a_+(x^+) \bigg )\;.
\end{split}
\end{align}

As in our preceding work~\cite{Barata:2022wim}, here we also take McLerran-Venugopalan (MV) model~\cite{McLerran:1993ka,McLerran:1993ni} to describe the field $\mathcal{A}$ that accounts for the background medium and we make use of high energy (eikonal) limit, where $p^+\gg |\p|\equiv p^\perp,p^-$.
%The medium is treated as a classical stochastic field, and the underlying approximation is the eikonal picture of $p^+>>p^\perp,p^-$, in other words, small angle deflection of a high energy projectile traversing a medium. 
Note, nevertheless, the full Hamiltonian method allows us to go beyond the formal eikonal limit of $p^+=\infty$. 
%In the basic picture, the color charges are the valence quarks, which then radiate gluons, represented by the classical field. 
The color charge density of the medium is assumed to have a Gaussian and local correlation function
\begin{multline}\label{eq:MV_color_charge}
 \llangle \rho_a(x^+,\x)\rho_b(y^+,\y) \rrangle \\
 =g^2 \mu^2\delta_{ab}\,\delta^{(2)}(\x-\y)\,\delta(x^+-y^+)\;,
\end{multline}
where we use $\llangle \cdots \rrangle$ to denote the average over medium configurations, and $\mu$ controls the strength of the medium. 
The saturation scale is defined as 
\begin{align}
 Q_s^2\equiv \frac{C_F g^4 \mu^2 L_{\eta}}{2\pi} \, ,   
\end{align}
with the fundamental Casimir $C_F= (N_c^2-1)/{(2N_c)}$.
%Numerically, the charges can be sampled as stochastic Gaussian variables. 
The field is then solved from the reduced classical Yang-Mills equation,
\begin{align}\label{eq:MV}
 (m_g^2-\nabla^2_\perp )  \mathcal{A}^-_a(x^+,\x)=\rho_a(x^+,\x)\;,
\end{align}
where the gluon mass $m_g$ is introduced to regularize the infrared (IR) divergence in the field \cite{Krasnitz:2001qu}. 

The time evolution of the quark jet, as a quantum state, obeys the time-dependent Schr\"{o}dinger equation.
Written in the form of path-ordered exponential, it reads,
% The time evolution of the probe, in the Schr\"{o}dinger picture, is controlled by the time evolution operator $U$,  such that the 
\begin{align}~\label{eq:evolution_eq}
\begin{split}
 \ket{\psi(x^+)}=& U(x^+;0)\ket{\psi(0)}\\
 \equiv& \cT_+ e^{-\frac{i}{2}\int_{0}^{x^+} \diff z^+\, P^-(z^+) } \ket{\psi(0)} \;,
\end{split}
\end{align}
where $\cT_{+}$ is the light-front time ordering operator, and $\ket{\psi(x^+)}$ the quantum state of the jet at time $x^+$.\footnote{Note that the factor of $\frac{1}{2}$ comes from the convention of the metric we are using: $x^+$ is conjugate to $p_+=\frac{1}{2}p^-$. This factor was mistyped in Equation (6) of our preceding paper Ref.~\cite{Barata:2022wim}. }
We solve this equation non-perturbatively by decomposing the time-evolution operator as a sequence of small time steps in the light-front time $x^+$,
\begin{align}~\label{eq:evolution_product}
 U(L_\eta;0)
=& \prod_{k=1}^{N_t} U(x^+_k;x^+_{k-1})\;,
\end{align}
where $x^+_k=k\, L_\eta/N_t$ is the intermediate time and $N_t$ the total number of time steps.

\subsection{Quantum simulation algorithm}
\label{sec:to_qcomputer}

The digital quantum simulation algorithm~\cite{Feynman:1981tf, Zalka:1996st, Wiesner:1996xg, nielsen_chuang_2010, Georgescu:2013oza} typically involves five generic steps: input, encoding, initial state preparation, time evolution, and measurement.
Here, we extend the algorithm developed in Ref.~\cite{Barata:2022wim} for a jet in the $\ket{q}$ Fock space to the $\ket{q}+\ket{qg}$ sectors.

\subsubsection{Basis encoding}\label{sec:basis_encoding}
We choose the eigenstates of the kinetic energy part of the Hamiltonian $P^-_{KE}$ as the basis states, as formulated in the Ref.~\cite{Li:2021zaw} for $\ket{q}+\ket{qg}$.
This basis choice is convenient in studying the momentum broadening of the jet state.  

We start by considering a generically truncated Fock space of the quark jet state.
The full Hilbert space of this theory can be formally decomposed as a tensor product over all single particle subspaces~\cite{Barata:2020jtq,Mueller:2019qqj}. Each Fock sector can have a finite projection in each one of these subspaces. 
Let us consider a generic multi-particle Fock sector in the quark jet state, $\ket{ q\ldots q \bar q \ldots \bar q g\ldots g}$, in which the number of quarks is one more than that of the anti-quarks.
The basis state is in the form of $\ket{\beta_{q\ldots q \bar q \ldots \bar q g\ldots g}}=\ket{\beta_q}\otimes \ldots \otimes\ket{\beta'_q}\otimes \ket{\beta_{\bar q}}\otimes\ldots\otimes \ket{\beta'_{\bar q}}\otimes\ket{\beta_g} \ldots \otimes\ket{\beta'_g}$.
Each single particle state carries five quantum numbers
\begin{align}\label{eq:basis_beta}
 \beta_l = \{p^+_l, p^x_l, p^y_l, c_l, \lambda_l \}, \text{ with } l=q,\bar q, g \;,
\end{align}
where $p^+$ is the longitudinal momentum, $\{p^x, p^y\}$ the transverse momenta, $\lambda$ the light-front helicity, and $c$ the color index. 
For a basis state in the truncated Fock space with up to $n+1$ quarks, $n$ anti-quarks and $m$ gluons,
\begin{align}\label{eq:multi_encode}
\begin{split}
    &\ket{\beta_{
    \underbrace{q\ldots q }_{n+1}
     \underbrace{\bar q \ldots \bar q}_n
     \underbrace{g\ldots g}_m}}\\
    &\quad \to 
    \ket{\beta_{q_0}} 
    \otimes \prod_{i=1}^{N_q} \Big(\ket{e_{q_i}}\otimes \ket{\beta_{q}^i} \Big)\\
    &\quad\quad
    \otimes \prod_{j=1}^{N_{\bar{q}}} \Big(\ket{e_{\bar q_j}}\otimes \ket{\beta_{\bar q}^j} \Big)
    \otimes 
    \prod_{k=1}^{N_g} \Big(\ket{e_{g_k}}\otimes \ket{\beta_{g}^k} \Big)\;,
\end{split}
\end{align}
in which $N_q$ is the total number of single quark basis states, $N_{\bar q}$ for antiquarks, and $N_g$ for gluons.
Each register $\ket{e_{q_i}}$, $\ket{e_{\bar{q}_i}}$, $\ket{e_{g_i}}$ encodes the occupancy of quarks, antiquarks, and gluons in the $i$-th single-quark basis state $\beta_{q}^i$, and they satisfy $\sum_i^{N_q} e_{q_i}=\sum_j^{N_{\bar q}} e_{\bar q_j}=n $ and $\sum_k^{N_g} e_{g_k}=m$. 
Particle exchange symmetry should be satisfied accordingly and implemented on the physical state; see discussion and references in~\cite{Barata:2020jtq} for further details.
The encoding of the single particle basis $\ket{\beta_l}$ can follow the strategy described in our previous work~\cite{Barata:2022wim}, which simply enumerates all the quantum numbers in the phase basis. 

Following this construction, let us describe in detail the encoding for the Fock space $\ket{q}+\ket{qg}$. 
We can save on the number of qubits with the following arrangement and simplification.
According to the strategy sketched in Eq.~\eqref{eq:multi_encode}, we need one qubit to encode the occupancy status of the gluon, e.g., $\ket{e_g}=\ket{0}$ for $\ket{q}$ and $\ket{e_g}=\ket{1}$ for $\ket{qg}$. 
We extend $\ket{e_g}$ to multi-qubits to also encode the $p^+$ quantum number of the gluon, denoted as $\ket{\zeta}$. 
In the helicity space, we make a simplification by considering only the helicity-non-flip term, i.e., $\lambda_q=\lambda_g=\uparrow$.
Note that the quark is taken to be massless, so only the quark-helicity-non-flip terms in $V_{qg}$ are non-zero, then the chosen configuration is the dominant contribution when the emitted gluon is soft. We therefore do not need extra quantum registers for the helicity space.
The remaining quantum numbers to be encoded are the transverse momenta and the color indices.
Therefore, the complete basis encoding for any basis state in the $\ket{q}$ and $\ket{qg}$ Fock sectors is written as
\begin{align}\label{eq:encoding_complete}
    \ket{\beta_{\psi}} \to \ket{\zeta}\otimes \underbrace{\Big(\ket{p_g^x} \ket{p_g^y}\ket{c_g}\Big)}_{\ket{g}} \otimes \underbrace{\Big(\ket{p_q^x} \ket{p_q^y}\ket{c_q}\Big)}_{\ket{q}}.
\end{align}
In the following, we recapitulate the encoding scheme for the transverse momentum and extend that for the color sector following our previous work~\cite{Barata:2022wim}, and elaborate on the construction of the $\ket{\zeta}$ register.

\begin{enumerate}
\item[i.] The transverse dimension

We formulate the transverse space as a two-dimensional square lattice. Both lattices span a size of $2 L_\perp$ and a number of $2 N_\perp$ sites per dimension such that the lattice spacing is $a_\perp=L_\perp/N_\perp$. We impose periodic boundary conditions on the lattice, such that this position space and the reciprocal momentum space are related by a discrete Fourier Transform, which on the quantum computer can be implemented via a quantum Fourier Transform ($q\mathcal{FT}$).
An arbitrary momentum state vector $\ket{\p} =\ket{p_x, p_y}$ is represented by a lattice coordinate $\ket{\k}=\ket{k_x, k_y}$ with $\p = \k\, b_\perp$ and $b_\perp = \pi/L_\perp$.
Similarly, we map any position vector $\ket{\x}=\ket{x_x, x_y}$ to the lattice position vector $\ket{\n}=\ket{n_x, n_y}$ with $\x = \n \, a_\perp$.

\item[ii.] The color space

We consider $N_c=2$, such that there are $N_c=2$ color degrees of freedom for the quark and $N_c^2-1=3$ for the gluon. 
We use one (two) qubit (s) to encode the color index of the quark (gluon). The quantum registers for the color space are explicitly, 
\begin{align}\label{eq:encoding_color}
\begin{split}
    \ket{c_q} &: \{ 0 \rightarrow \ket{0} ,\quad 1\rightarrow \ket{1}\}\;,\\
    \ket{c_g} &: \{ 0 \rightarrow \ket{00},\quad 1 \rightarrow \ket{01} ,\quad 2 \rightarrow \ket{10}\}\;.
\end{split}
\end{align}

\item[iii.] The $\ket{\zeta}$ register

We compactify $x^-$ to a circle of length $2L$ (i.e., $x_+$ to a circle of length $L$), and impose periodic boundary conditions for bosons and anti-periodic conditions for fermions, such that the longitudinal momentum $p^+$ is discretized,
\begin{align}\label{eq:basis_long}
\begin{split}
    & p^+_l=\frac{2\pi}{L}k^+_l, \\
    & k^+_q=\frac{1}{2},\frac{3}{2},\cdots,\quad
    k^+_g=1,2,3,\cdots
\end{split}
\end{align}
where the zero mode for the gluon is excluded.\footnote{The $p^+=0$ zero modes on the light front are usually complicated, and not relevant for our problem. Refer to ~\cite{Yamawaki:1998cy, Brodsky:1997de} for a review.} 

The total $p^+$ is preserved by the Hamiltonian. 
Consider a quark jet with a definite $P^+$, then each basis state has the same total longitudinal momentum, i.e., $\sum_i p^+_i=P^+$ in which $i$ enumerates the Fock particles. 
As such, in the $\ket{q}$ sector, $p^+_Q\equiv P^+$; in the $\ket{qg}$ sector, $p^+_q+p^+_g\equiv P^+$.
\footnote{Here and throughout the paper, we use the subscripts “Q” and “q” to distinguish between the quark in the $\ket{q}$ sector and that in the $\ket{qg}$ sector.}
We introduce the total longitudinal quanta $K$ such that
\begin{align}\label{eq:basis_long_2}
K\equiv\sum_i k^+_i=k_q^++k_g^+, \quad P^+=\frac{2\pi}{L}K,
\end{align}
where $K$ is a positive half integer. 

We combine the longitudinal encoding with the gluon occupancy using the quantum register $\ket{\zeta}$. 
The $\zeta=0$ state on the register encodes the $\ket{q}$ state with $k_Q^+ = K$; and $\zeta=\{1,2,\cdots K-1/2\}$ encodes the $\ket{qg}$ states with $k_g^+=\{1,2,\cdots,K-1/2\}$ and $k_q^+=K-k_g^+$. 
In the latter case, the value of $\zeta$ relates to the gluon's longitudinal momentum fraction as $z_g\equiv p^+_g/P^+=\zeta/K$.
Written explicitly,
\begin{align}\label{eq:encoding_long}
   \zeta \rightarrow (k^+_g, k^+_q) =
    \begin{cases}
        0 \rightarrow  (*, K) \\
         1 \rightarrow  (1, K-1)\\
         2 \rightarrow  (2, K-2)\\
         \quad\vdots \\
         K-\frac{1}{2} \rightarrow (K-\frac{1}{2}, \frac{1}{2})
    \end{cases}\;,
\end{align}
where we use $*$ to represent the absence of the gluon in the $\ket{q}$ sector.

\end{enumerate}

This basis space contains a total number of $2^3\, \Kceil\, (2 N_\perp)^4$ basis states, which scales as a power four with the lattice size. On a classical computer, the problem quickly deteriorates when additional Fock sectors are included, requiring at least ($2N_\perp)^{2 n}$ resources for $n$ particles. On the quantum circuit, however, we would only need a total of $n_Q=(7+4\log_2{N_\perp}+\log_2{\Kceil})$ qubits for this problem, dramatically reducing the number of resources. Nevertheless, efficient gate simulation is still needed and not necessarily always guaranteed.

Using this qubit encoding scheme encapsulated by~\eqn{eq:encoding_complete}, one can prepare any initial state $\ket{\psi_0}$ as a superposition of basis states. Though arbitrary state initialization may be difficult, the preparation of many useful choices of initial states is feasible~\cite{Deliyannis:2021che,KitaevGauss}. Since we are mostly interested in studying the jet evolution in momentum space, we neglect the initial state effects, and take the initial state $\ket{\psi_0}$ to be a zero transverse momentum and even-color single-quark state, unless specified otherwise.

\subsubsection{Gate encoding and time evolution}\label{sec:gate_encoding}

\begin{figure}
    \centering
    \subfigure[\, Schematic representation of the digital circuit 
    ]{\includegraphics[width=0.45\textwidth]{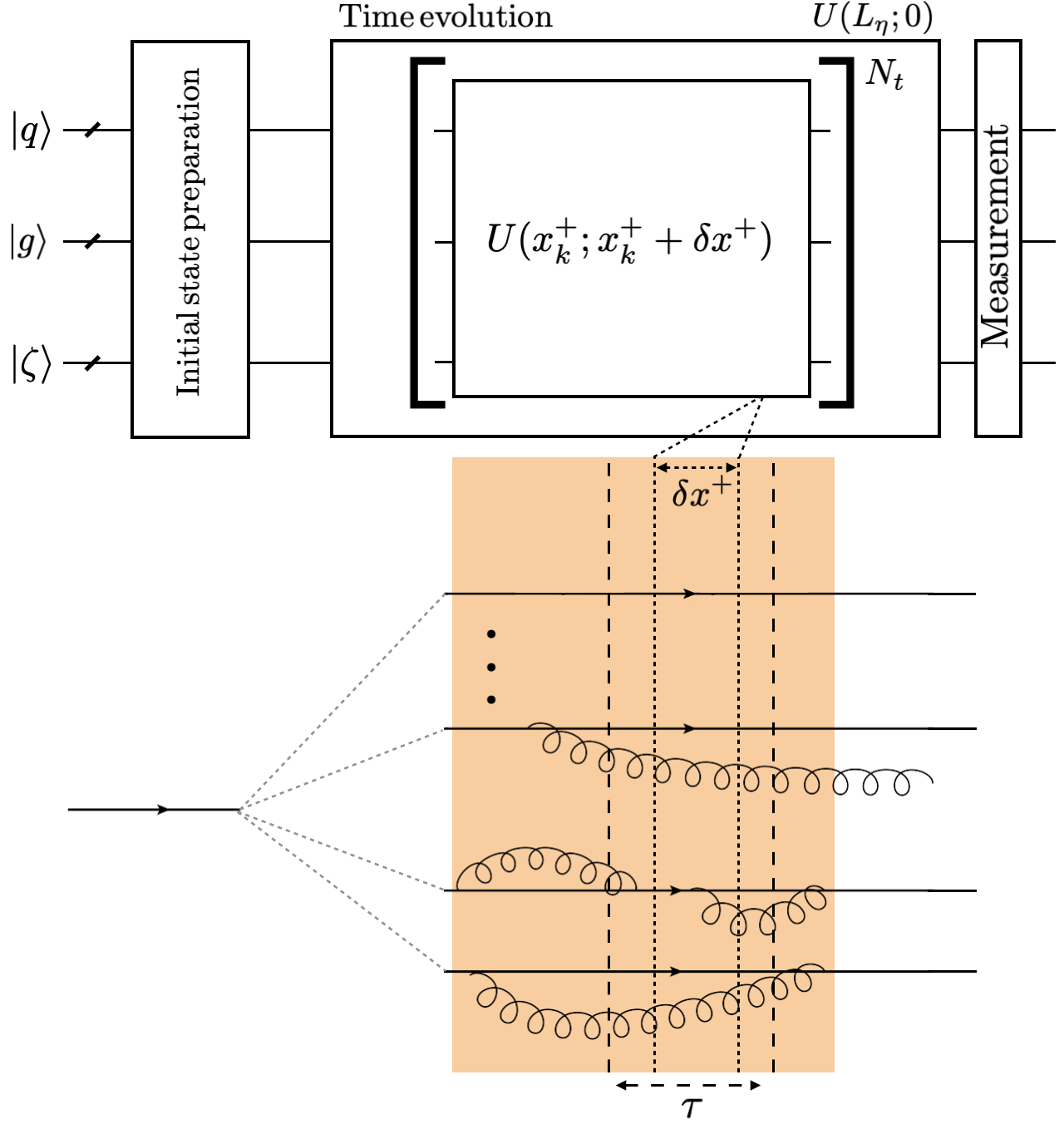}}
    %\vspace{2 cm}
    \subfigure[\,Two treatments of $U(x_k^+; x_k^+ +\delta x^+)$]{ \includegraphics[width=0.45\textwidth]{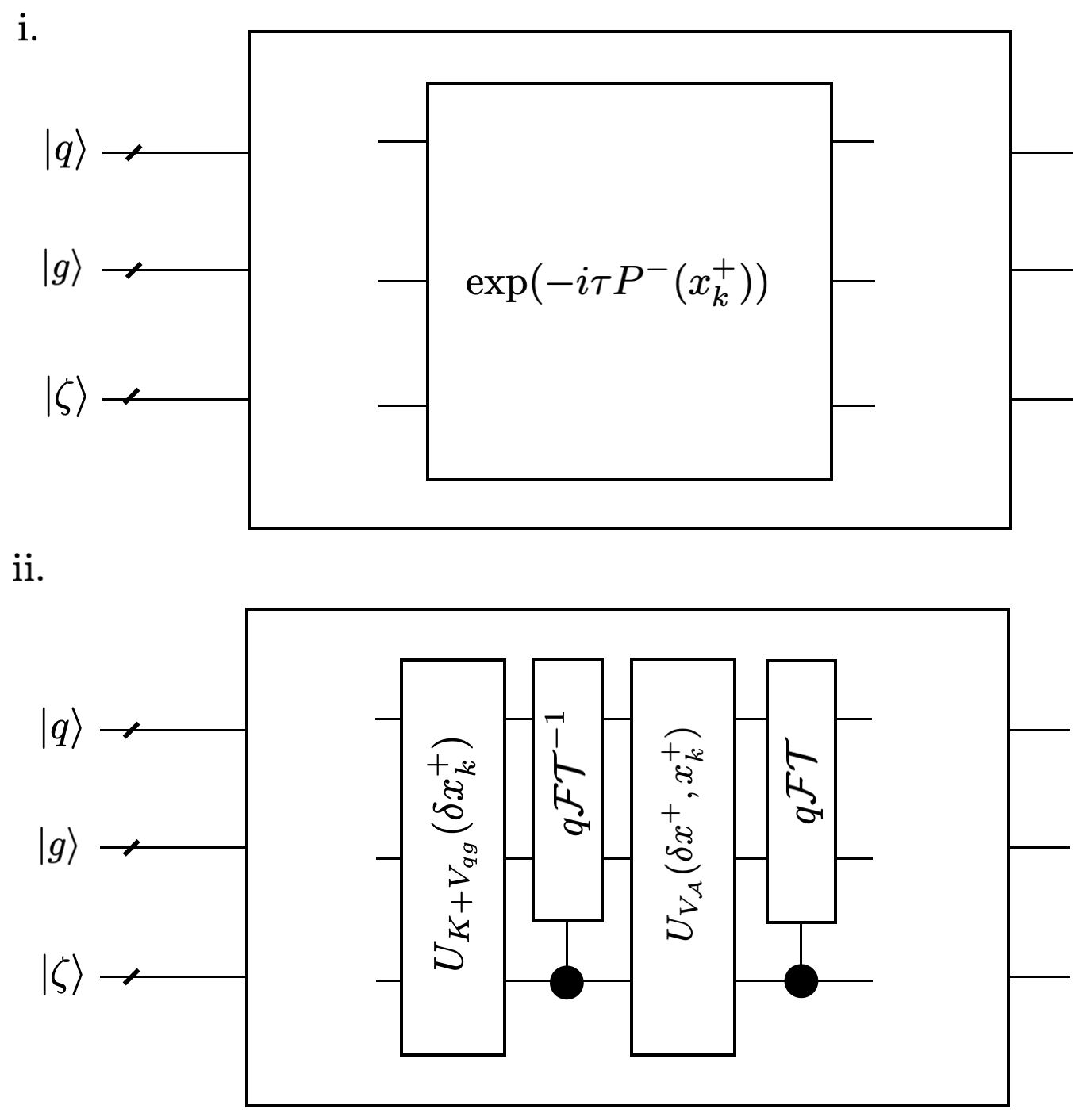}}
    \caption{Schematic representation of the digital circuit used to simulate the multi-parton jet in the medium.
    (a) The top panel is a quantum circuit of the whole simulation process with registers for the quark, the gluon, and the occupancy status.
    The bottom panel illustrates that the jet is a superposition of all possible quantum states in the phase space, and the medium shown in the yellow band expands through the process.
    The single timestep evolution block in the circuit corresponds to a time slice of $\delta x^+$ in the whole process, and $\tau$ is the duration of each medium layer.
    (b) Two treatments of the evolution operator in a single timestep: i. direct exponentiation, and ii. 
    alternating exponentiation in a momentum-position-mixed space. See more discussions in the text.
    }
    \label{fig:sketch}
\end{figure}

We implement the product formula decomposition by splitting the evolution along the $x^+$ direction into $N_t$ time steps, each with a duration of $\delta x^+=L_\eta/N_t$, as in Eq.~\eqref{eq:evolution_product}. 
Note that the time-dependence of the Hamiltonian is from the background field $\mathcal A$. 
We slice the medium into $N_\eta$ layers along $x^+$~\cite{Lappi:2007ku,Ipp:2020mjc,Li:2020uhl,Barata:2022wim, Li:2021zaw, Li:2023jeh}, such that the time duration for each layer is $\tau\equiv L_\eta/N_\eta$.
A schematic representation of the circuit is presented in Fig.~\ref{fig:sketch}. We generate the values of the background field $\mathcal{A}$ beforehand in a classical computer, as in our preceding work~\cite{ Barata:2022wim}. The details on how to calculate the field numerically are given in App.~\ref{app:field}. 
 
We consider and compare two treatments on the evolution operator.
\begin{enumerate}
\item[i.] Direct exponentiation 

For a Hamiltonian that is constant in time, the evolution
operator reduces to an ordinary exponential, which can be evaluated directly. 
This is the case for our Hamiltonian within each layer of the medium.
Taking $\tau$ as the size of the time step $\delta x^+$, such that $N_t=N_\eta$, we therefore have the single-step evolution operator as
\begin{align}\label{eq:unitary_mom}
\begin{split}
 U(x_k^++\tau; x^+_k) =\exp \left\{-i \tau P^-(x_k^+)\right \}
    \;,
\end{split}
\end{align}
in which $k=1,2,\ldots, N_\eta$.
Here, $P^- =K+V_{qg}+V_{\mathcal{A}}$ denote the matrix form 
of the corresponding operator evaluated on the basis space.\footnote{Full
expressions of the matrix elements are found in Refs.~\cite{Li:2020uhl, Li:2021zaw}. } 
We evaluate the Hamiltonian matrix elements in the transverse momentum basis space, such that the K part is diagonal and the $V_{qg}$ part is off-diagonal and sparse. However, the $V_{\mathcal{A}}$ term is more complex since it has scattered elements in momentum space. Alternatively, one can perform the equivalent calculation in the transverse position basis, which favors the evaluation of $V_{\mathcal{A}}$ but complicates the evaluations of $K$ and $V_{qg}$. 

The major advantage of this treatment is that the evaluation is exact up to numerical accuracy.
Nonetheless, the disadvantage of this treatment is the large time complexity of obtaining the Pauli strings of these respective non-diagonal operators; though for small-sized problems, it is feasible to take this treatment.

\item[ii.] Alternating exponentiation (mixed-space simulation)

When the time step is sufficiently small, the Hamiltonian within a single step can be considered as constant, and one can further ``trotterize" the single-step evolution operator as a product of operations with the different  components of the Hamiltonian.
In this way, one can factorize the single-step evolution operator into a series of unitary operators from different components in the Hamiltonian and boost the computational efficiency, e.g., Refs.~\cite{Barata:2022wim, Li:2021zaw}.
We split the single-step evolution operator as the following,
\begin{align}\label{eq:unitary_mixed}
\begin{split}
   \quad  U(& x_k^++\delta x^+; x^+_k) \\
    \approx & [q\mathcal{FT}] \exp \bigg\{-i\delta x^+ \left[{V}_{\mathcal{A}}(x_k^+)\right]\bigg \} [q\mathcal{FT}^{-1}]\\
    \times&\exp \bigg\{-i\delta x^+ \left[ {K}+ {V}_{qg} \right]\bigg \} ,
    \end{split}
\end{align}
in which $k=1,2,\ldots, N_t$. Note that here $\delta x^+\leq \tau$, i.e., $N_t\geq N_\eta$.
In practice and in our simulations, the time step $\delta x^+$ is taken sufficiently small (i.e., $N_t$ sufficiently large) to ensure the result is convergent when comparing to smaller $\delta x^+$s; discussions on the convergence of $N_t$ and $N_\eta$ can be found in App.~\ref{app:NtNeta_converge}.

The evolution of the kinetic energy and gluon emission/absorption, as the operation with ${K}+{V}_{qg}$, is performed in the momentum space; whereas the evolution with the medium, that with ${V}_{\mathcal{A}}$, is performed in the position space. 
Since the background field $\mathcal{A}(x_k^+, \x)$ is diagonal in the transverse position space, this mixed-space evolution approach is the most economical way of evaluating the Pauli terms and could potentially extend to larger lattice and more Fock spaces. Detailed resource cost comparison between the Pauli term evaluations in momentum/position space is included in App.~\ref{app:Pauli_strings}. The basis transformation between the momentum and position spaces can be performed by applying a quantum Fourier transform $q\mathcal{FT}$ to $\{\ket{\n}\} \rightarrow \{\ket{\k}\}$, and its inverse $q\mathcal{FT}^{-1}$ to $\{\ket{\k}\} \rightarrow \{\ket{\n}\}$ per each transverse dimension of the quark and the gluon. 

\end{enumerate}

To perform the simulation using either of the aforementioned treatments, one needs to implement the quantum gates of an operation in the format of $e^{iH\delta x^+}$ (with $M$ a Hermitian operator). In our previous work \cite{Barata:2022wim}, the matrix elements of $e^{iH\delta x^+}$ can be obtained exactly using the properties of the exponential of the Pauli vector and then transcribed to unitary gates using the quantum Shannon Decomposition~\cite{shende2006}. Though this approach works well for simulating the jet in the quark Fock space, it is inconvenient to obtain the exact exponential of the type of Hamiltonian in this work. 

Instead, we find the corresponding Pauli terms of $H$ first and then the associated quantum gates, since there is a direct correspondence between the Pauli exponentials and the quantum gates~\cite{nielsen_chuang_2010}. To obtain the Pauli terms, various strategies can be adopted (see App.~\ref{app:Pauli_strings} for examples and discussions), and we used the sparse matrix projection methods to take advantage of the property of the Hamiltonian matrix. To time evolve our Pauli terms, we use {\tt PauliEvolutionGate} class provided by {\tt Qiskit}~\cite{Qiskit}, which automatically maps the Pauli operators to quantum gates.
For small problem sizes, we can perform exact operator evolution via matrix exponentiation; for large problem sizes, we can use the Lie-Trotter formula~\cite{Trotter:1959} to approximate the exponential of non-commuting operators at first order. For higher-order approximations, we can use the Suzuki-Trotter product formula~\cite{Suzuki:1976}. All these methods are conveniently implemented in various {\tt Synthesis} classes~\cite{nielsen_chuang_2010, Hatano:2005gh} in {\tt Qiskit}. We studied the performance of these unitary exponential implementations, especially using {\tt MatrixExponential}, {\tt LieTrotter}, and {\tt SuzukiTrotter}, and found that their performances are almost identical with each other at our problem scale.
 
\subsubsection{Measurement}\label{sec:method_measure}

We extract the information about the final quantum state by directly measuring the prepared state. In practice, since we work in small lattice sizes, such an approach is the most efficient. Note however, it is not always necessary to measure the full quantum state. For example, to obtain the induced gluon probability, one can measure the $\ket{\zeta}$ quantum register on a $\log_2(K)$-bit classical register alone, greatly reducing the number of measurement shots needed. 

While most of the results presented in this work use the shot-based {\tt QasmSimulator} backend to extract physical observables such as momentum broadening, we also use the {\tt StatevectorSimulator} backend to capture the exact quantum state, serving as a benchmark. Though not practical on real quantum devices, it allows us to study the information flow in the evolution of the quark jet in medium. Estimating the entropy directly on the quantum circuit is generally difficult~\cite{Subramanian:2021, Acharya:2020, TongyangLi:2019}, and requires full-fledged fault-tolerant quantum computers in the future. 

%%%%%%%%%%%%%%%%%%%
\section{ Quantum simulation results}
\label{sec:results}
%%%%%%%%%%%%%%%%%%%%%

In this section, we study the quantum simulation results for the evolution of a jet in a dense stochastic medium, using the light-front Hamiltonian formalism and quantum simulation method introduced in the preceding sections. 
Specifically, we focus on the momentum broadening of the jet, the gluon emission, and the entropy growth, for several backgrounds with different medium strengths. 
We perform the simulations using the ideal {\tt QASM} simulators from \texttt{Qiskit}.

For the simulations, we take the transverse lattice with $N_\perp=1$ for the $\ket{q}+\ket{qg}$ system (we will also use larger $N_\perp$ when examining the $\ket{q}$ system), and the total longitudinal momentum quanta $K=3.5$. Although these numbers are small, it still provides us with a two-by-two transverse lattice for both the quark and gluon single particle states, allowing investigation on the effects of momentum broadening. 
With $K=3.5$, we are also able to examine the distribution of the longitudinal momentum. We take $L_{\perp}=32~\GeV^{-1}= 12.6$~fm. The duration of the medium is taken to be $L_{\eta}=50\GeV^{-1}$= 9.87 fm.
We take the layer number to be $N_\eta=4$; one can find the discussions on the convergence of $N_\eta$ and the evolution time steps $N_t$ in Appendix.~\ref{app:NtNeta_converge}.
The IR regulator for the medium is $m_g=0.8 \GeV$.  More details on the determination of parameters for a proper lattice and medium can be found in our previous works~\cite{Li:2021zaw, Barata:2022wim}. The total number of qubits required in this setup is therefore $n_Q = 9$ according to the encoding scheme in Sec.~\ref{sec:basis_encoding}. 

Since we are mostly using the shot-based quantum simulator, we make sure a sufficient number of counts are used to sample the true probability distribution. Unless stated otherwise, we always use 819200 shots, which proves to be more than enough to take into account of the noise from statistic sampling for a 9-qubit simulation~\cite{Barata:2022wim}. The uncertainties (i.e. standard deviations) provided on our plots are therefore exclusively related to the medium field fluctuations arising from using a stochastic medium in the MV model. With these in mind, we will present our main results in the following.

%%%%%%%%%%%%%%%%%%
\subsection{Momentum broadening}
%%%%%%%%%%%%%%%%%%

Transverse momentum broadening is an important observable to understand the evolution of the jet inside the medium.
We examine the square of the transferred momentum $\Delta\braket{p^2_\perp ( \Delta x^+)}$ at various medium strengths of $g^2\mu$, which simplifies to $\braket{p^2_\perp ( \Delta x^+)}$ when the initial state has a zero transverse momentum. 

In the eikonal limit, $\braket{p^2_\perp ( x^+)}$ of a single particle is linear in time, and the proportionality constant can be interpreted as the quenching parameter $\hat q$. We have provided the explicit expression in the chosen basis representation in our previous work, as in Eq.(19) of Ref.~\cite{Barata:2022wim}.\footnote{
Here, we write out the analytical expectation at the special case of $N_\perp=1$ in order to compare with the simulation results. 
The specialty of the phase space at $N_\perp=1$ is that the lattice UV and IR cutoffs estimated in the usual way as $\lambda_{UV}=\pi/a_\perp$ and $\lambda_{TR}=\pi/L_\perp$ would be the same, then the analytical formula for general $N_\perp$ no longer hold. One should instead, treat the $\vec p_\perp$ integral as a sum over the full discrete space of $\vec p_\perp$.
In this way, we get
\begin{align}\label{eq:qhat_x_lattice_N1}
 \begin{split}
  \Delta& \braket{p^2_\perp ( x^+,  x=a_\perp  m_g/\pi, N_\perp=1))} \big|_{\text{on lattice}}\\
  &  =  C_F
   g^4 \tilde{\mu}^2 
   \frac{1}{(2 \pi)^2} 
   \left[
   \frac{2 }{ (x^2 +1)^2}
   +    \frac{2 }{ (x^2 +2)^2}
   \right] \Delta x^+
   \;.
 \end{split}
\end{align}
In analogy, the $\braket{p^2_\perp ( x^+)}$ for a gluon state replaces $C_F=(N_c^2-1)/(2 N_c)$ by $C_A=N_c$ in the above equation.
For an uncorrelated quark-gluon state, one should replace $C_F$ by $C_F+C_A$ in the above equation.
}
We will use the eikonal expectation to verify our simulation results in the eikonal limit and examine non-eikonal effects by studying the deviation from them.

With the final jet probability distribution extracted from the quantum simulation, we are able to reconstruct the total transverse momentum of the jet, given as
\begin{align}\label{eq:p2_def}
    \begin{split}
\braket{p^2_\perp}
&= \braket{\psi(L_\eta)| \hat{p}^2_\perp|\psi(L_\eta)}\\
&=\mathcal{P}_{\ket{q}}\braket{p^2_{\perp}}_{\ket{q}} + \mathcal{P}_{\ket{qg}}\braket{p^2_{\perp}}_{\ket{qg}} \;,
    \end{split}
\end{align}

where $\mathcal{P}_{\ket{q}}$ ($\mathcal{P}_{\ket{qg}}$) is the probability of the state in the quark (quark-gluon) Fock sector. 
With a zero momentum initial state, $\braket{p_\perp^2}$ indicates the broadening effect exclusively due to the medium. By comparison, $\braket{p_\perp^2}=0$ in vacuum due to momentum conservation. 
The periodic boundary conditions of the lattice are taken into account when summing the momenta of the quark and gluon state; see the prescription in Appendix C of Ref.~\cite{Li:2021zaw}.

\begin{figure}[!t]
    \centering
    \includegraphics[width=0.5\textwidth]{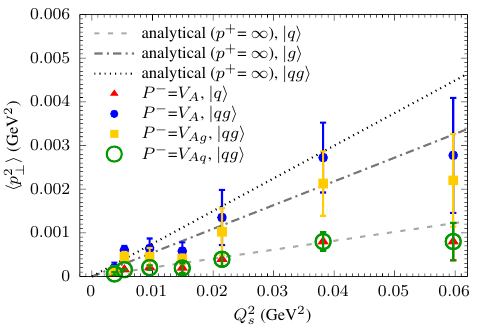}
    \caption{The dependence of the momentum broadening $\braket{p_\perp^2}$ on the saturation scale $Q_s$, for various particles. 
    The Hamiltonian contains only the medium interaction, as specified in the legends.
    The eikonal analytical results in the dashed and dotted lines are given by Eq.~\eqref{eq:qhat_x_lattice_N1}.
    }
    \label{fig:p2_sub}
\end{figure}

We first verify our method by keeping only the medium interaction term in the Hamiltonian, $V_{\mathcal{A}} = V_{q\mathcal{A}} + V_{g\mathcal{A}}$.
This setup corresponds to the process in the eikonal limit of $p^+=\infty$, therefore the expectation value $\braket{p^2}$ should agree with the eikonal expectation, e.g., Eq.~\eqref{eq:qhat_x_lattice_N1}. 
Specifically, we assign the initial state as both a single quark and a single quark-gluon state with total transverse momentum $\p=\mathrm{0}$,\footnote{Note that for the quark-gluon initial state, we take $\p_q=\p_g=\mathrm{0}$. 
This means that in the transverse position space, the two particles are maximally delocalized, so their correlation is negligible; the exact correlation relation is derived and given in Ref.~\cite{Li:2023jeh}.
For this reason, we use the uncorrelated quark-gluon analytical result, as given by Eq.~\eqref{eq:qhat_x_lattice_N1} with $C_A+C_F$ as the Casimir, as its eikonal reference. 
} and we put in the Hamiltonian $V_{q\mathcal{A}}$ and $ V_{g\mathcal{A}}$ separately and in-combined. 
We present in Fig.~\ref{fig:p2_sub} the results of the final state $\braket{p^2}$ at various saturation scales $Q_s$.
The obtained simulation results agree with the expected eikonal analytical results. 
Similar to the single quark results shown in our previous work~\cite{Barata:2022wim}, the $\braket{p^2}$ exhibits increased uncertainty at larger saturation scales, which is related to the larger Gaussian width in constructing the stochastic background fields; see also on the background field in App.~\ref{app:field}. In addition, the $\braket{p^2}$ starts to bend as the saturation scale increases, as a result of the lattice admitting a UV cutoff of $\pi/a_\perp$.

\begin{figure}[!t]
\centering
    \subfigure[\label{fig:p2_Pplus1} $p^+=1$ GeV]{
    \includegraphics[width=0.5\textwidth]{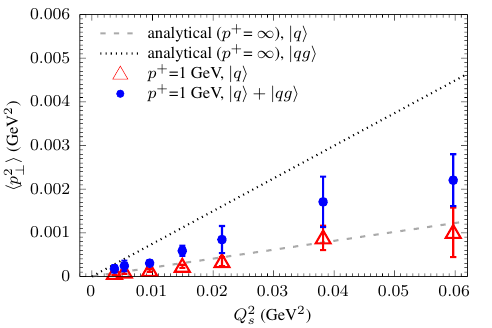}}
    \quad
    \subfigure[\label{fig:p2_Pplus1000} $p^+=1000$ GeV]{
    \includegraphics[width=0.5\textwidth]{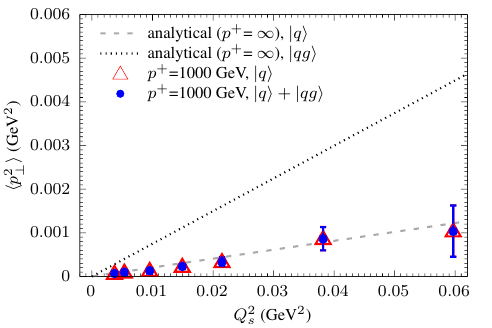}}
    \caption{The dependence of the momentum broadening $\braket{p_\perp^2}$ on the saturation scale $Q_s$, at (a) $p^+=1$ GeV and (b) $p^+=1000$ GeV. 
    The initial state is a bare quark with $\vec p_\perp=\vec 0_\perp$.
    The results of the simulation in the $\ket{q}$ Fock space is in the open triangle, and that in the $\ket{q}+\ket{qg}$ Fock space is in the disk.
    The eikonal analytical results in the dashed and dotted lines are given by Eq.~\eqref{eq:qhat_x_lattice_N1}.
    }
    \label{fig:p2_full}
    
\end{figure}

We then perform the simulations with the full Hamiltonian $P^-=K+V_{qg}+V_{\mathcal{A}}$. 
The eikonal approximation is relaxed by letting the jet state have finite energy, $p^+=1, 1000$ GeV.
We assign the initial state as a single quark with $\p=0, k^+=K$, and use both the $\ket{q}$ and the $\ket{q}+\ket{qg}$ Fock space for the simulation. 
The results of the final state $\braket{p^2}$ at various saturation scales $Q_s$ is presented in Fig.~\ref{fig:p2_full}.
We have confirmed that the two simulation treatments discussed in Sec.~\ref{sec:gate_encoding}, the direct and the alternating exponentiation, led to the same results.
In the figure, the results in the eikonal analytical limit in the $\ket{q}$ and $\ket{qg}$ Fock spaces according to \eqn{eq:qhat_x_lattice_N1} are provided in the dashed and dotted lines for comparison. 
We find that the $p^+=1000$ GeV result overlaps with the eikonal limit for the $\ket{q}$ with uncertainties taken into account;\footnote{For a fair comparison, all the simulation results in Fig.~\ref{fig:p2_full} used the same sets of medium background fields.} this is because both the kinetic energy and the gluon emission contribution are highly suppressed at the near eikonal limit ($p^+=\infty$), and therefore the occupancy in the $\ket{qg}$ sector is negligible.\footnote{From the Hamiltonian matrix element(e.g., as in Ref.~\cite{Li:2021zaw}) point of view, the $P^+$ is on the denominator in the $V_{qg}$ term.} 
By contrast, the $p^+= 1 \GeV$ result in $\ket{q}+\ket{qg}$ lies between the two eikonal limits, whose deviation from the single quark's eikonal expectation indicates non-eikonal effects due to gluon emission.
A simple and intuitive understanding is that the inclusion of the $\ket{qg}$ sector enlarges the phase space, and as a result enhances the momentum broadening effect~\cite{Li:2021zaw}. 

%%%%%%%%%%%%%%%%%%
\subsection{Gluon production }
%%%%%%%%%%%%%%%%%%

\begin{figure}[!t]
    \centering
    \subfigure[\label{fig:gluon_prob_vacuum_g1} $g=1$]{
    \includegraphics[width=0.5\textwidth]{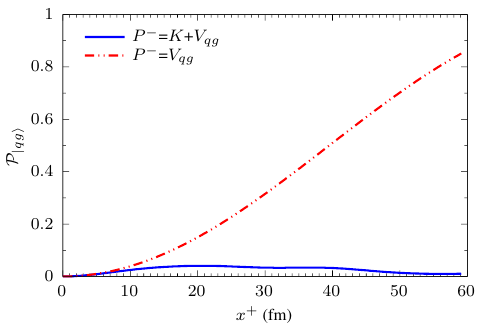}}
    \quad
    \subfigure[ \label{fig:gluon_prob_vacuum_g10} $g=10$]{
    \includegraphics[width=0.5\textwidth]{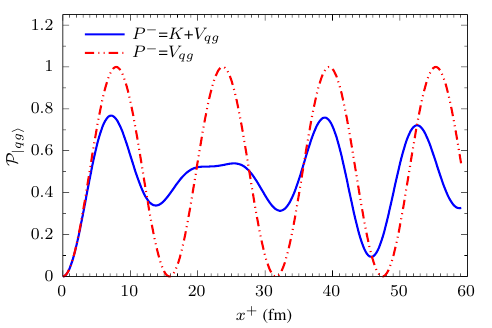}}
    \caption{Probability of the $\ket{qg}$ sector in the vacuum as a function of the evolution time $x^+$ with coupling strength (a) $g=1$ and (b) $g=10$ calculated using quantum simulators.
    The Hamiltonian contains the gluon emission term, with and without the kinetic energy term, as specified in the legends.
    Each curve is obtained from both quantum simulations and classical diagonalization, and the two are in agreement.
    }
    \label{fig:gluon_prob_vacuum}
\end{figure}

With the quark jet formulated as a superposition of $\ket{q}$ and $\ket{qg}$ states, it is interesting to study the gluon production through the evolution. 
In particular, we study the evolution of the probability of the jet in the $\ket{qg}$ sector, i.e., $\mathcal{P}_{\ket{qg}}$.
Furthermore, we examine the distribution of the gluon's longitudinal momentum fraction $z_g$.
We obtain the $z_g$ distribution
by performing projective measurement on the $\ket{\zeta}$ register.
For example, with $K=7/2$, we should have 4 different longitudinal modes across the $\ket{q}$ and $\ket{qg}$ Fock sectors:
\begin{align}
\begin{split}
        &\ket{q(k_q^+=7/2)},\\  &\ket{q(k_q^+=5/2)g(k_g^+=1)},\\ &\ket{q(k_q^+=3/2)g(k_g^+=2)},\\ &\ket{q(k_q^+=1/2)g(k_g^+=3)},
\end{split}
\end{align}
where the total longitudinal momentum quanta of each mode is always $K$. The possible longitudinal momentum fractions of the gluon can be read conveniently as $z_g = k_g^+/K = \{0.29, 0.57, 0.86\}$. 
To observe the probability distribution throughout the evolution time $L_\eta$ on the quantum simulator, the same simulation is repeated for different $x^+$ to extract the corresponding probability.

\subsubsection{Vacuum case}

To better compare the medium corrections, we first present the simulation of the initial quark jet in the vacuum, which can be achieved by simply turning off the $V_\mathrm{A}$ term in the Hamiltonian while keeping the $K$ and $V_{qg}$ terms.

In Fig.~\ref{fig:gluon_prob_vacuum}, we show the total probability of the $\ket{qg}$ Fock sector, $\mathcal{P}_{\ket{qg}}$, as a function of evolution time $x^+$ with and without the kinetic energy term, and using two different coupling strengths, $g=1$ and $g=10$.
The setup of having only the gluon emission term $V_{qg}$ (without $K$) in the Hamiltonian, though not physical, is important to help understand its effect.
For the size of the problem being simulated, it is also feasible to diagonalize the Hamiltonian and obtain the eigenstates, therefore knowing exactly the evolution of a given initial state. 
The quantum simulation results agree with the diagonalization results, which help verify our quantum simulation algorithm.
In both cases of $g=1$ and $10$, the $\ket{qg}$ probability oscillates periodically under just the $V_{qg}$ term, but this behavior is broken with the inclusion of the kinetic term, as expected. The effect of including the kinetic energy term is akin to the existence of an energy-dependent phase for different states, leading to a decoherence effect when summing over states. Similar behavior was seen in the classical simulations done in Ref.~\cite{Li:2021zaw}. 

The comparison between the results with $g=1$ and $g=10$ is also interesting.
In the pure $V_{qg}$ case, the oscillation frequency is proportional to $g$ and the amplitude is 1. 
In the $V_{qg}+K$ case, the amplitude of the decohered oscillation is much larger with the stronger coupling.
This is expected by noting that the oscillation amplitude is approximately proportional to the square of the ratio between the averaged $V_{qg}$ and the $K$ terms~\cite{Li:2021zaw}. 
We note that the oscillation amplitude is small at $g=1$ in Fig.~\ref{fig:gluon_prob_vacuum_g1} for the lattice that we are using. 
In principle, one can increase the lattice size $N_\perp$ to obtain larger oscillation amplitude; however, the simulation would be more expensive.
For the purpose of this work, we will present our results using the larger coupling strength $g=10$ in the $V_{qg}$ term unless specified otherwise. 

\begin{figure}[t]
    \centering
    \includegraphics[width=0.5\textwidth]{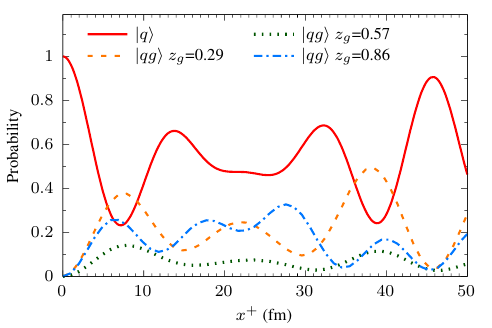}
    \caption{
    Evolution of the probabilities of different $p^+$ states, including the $\ket{q}$ sector and the different segments of the $\ket{qg}$ sector characterized by the gluon longitudinal momentum fraction $z_g$. 
    }   
\label{fig:gluon_prob_vacuum_g10_zg_style2}
\end{figure}

\begin{figure}[t]
    \centering
    %\hspace{-0.2in}
    \includegraphics[width=0.5\textwidth]{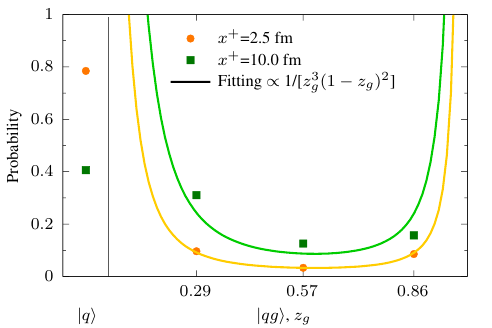}
    \caption{
    The probability of the quark state at different $p^+$ configurations (characterized by Fock sector and $z_g$) in the vacuum at selected time instances of $x^+=2.5, 10.0$ fm. 
    The solid lines are fits to the given functional form up to an overall constant.
    }
    \label{fig:gluon_prob_vacuum_g10_zg}
\end{figure}

In Fig.~\ref{fig:gluon_prob_vacuum_g10_zg_style2}, we present the evolution of the probabilities of different $p^+$ states, including the $\ket{q}$ sector and the different segments of the $\ket{qg}$ sector characterized by the gluon longitudinal momentum fraction $z_g$. 
The $\ket{qg}$ modes with the smallest $z_g$ dominate, as having a more rapid initial growth and a larger oscillation magnitude compared to the other two. 
Since we are simulating the spin non-flipping case, we expect the distribution of the longitudinal momentum fraction roughly proportional to the reduced splitting function $ P_{q\to qg}(z_g)\equiv 1/[z_g^3 (1-z_g)^2]$, according to the Hamiltonian matrix element.
Note that this splitting function matches the leading order $q\to q+g$ Altarelli-Parisi splitting function~\cite{Dokshitzer:1977sg, Altarelli:1977zs, Gribov:1972ri}, up to an integration measure which has to be included at the cross-section level. We numerically demonstrate a good agreement between data and the reduced splitting function $P_{q\to qg}$ in Fig.~\ref{fig:gluon_prob_vacuum_g10_zg}, up to a $x^+$ dependent state normalization constant. 

\subsubsection{Medium case}

\begin{figure}[t]
    \centering
    \includegraphics[width=0.50\textwidth]{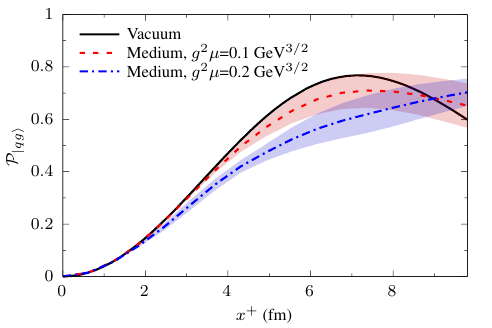}
    \caption{Probability of the $\ket{qg}$ component as a function of the evolution time $x^+$ with medium strength $g^2\mu =0.1, 0.2 \GeV^{3/2}$. 
    The result in vacuum is in the solid black line for comparison.
    }\label{fig:medium_evo}
\end{figure}

\begin{figure}[t]
    \centering
    \includegraphics[width=0.5\textwidth]{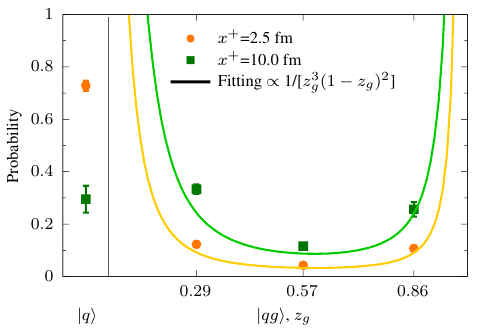}
    \caption{
    The probability of the quark state at different $p^+$ configurations (characterized by Fock sector and $z_g$) in medium with $g^2\mu =0.2 \GeV^{3/2}$ at selected time instances of $x^+=2.5, 10.0$ fm. 
    The solid lines are fits to the given functional form up to an overall constant.
    }
    \label{fig:medium_evo_by_z}
\end{figure}

The same observables can also be computed for the case of in-medium propagation and directly compared to the vacuum scenario, allowing us to visualize the modifications to the jet fragmentation pattern. In Fig.~\ref{fig:medium_evo}, we present the probability of the $\ket{qg}$ Fock states for an initial quark jet going through the colored mediums.\footnote{Quantum simulation with two different strategies in momentum-space and mixed-space are discussed in App.~\ref{app:NtNeta_converge} and their agreement is presented in Fig.~\ref{fig:convergence_Ndx}.}
Specifically, we used two sets of mediums with $g^2\mu=0.1~\mathrm{GeV}^{3/2}$ and $0.2~\mathrm{GeV}^{3/2}$.
Note $g=1$ in the medium term $V_{\mathcal{A}}$ whereas $g=10$ in the $V_{qg}$ term. 
Since the number of momentum modes is small, the decoherence among different modes is not sufficient to suppress the oscillation, even at late times. It is therefore hard to conclude whether the medium induces or suppresses the production of radiation conclusively. At this level, it is then not possible to fully comment on the relation between our numerical results and the Landau-Pomeranchuk-Migdal (LPM), which is known to determine the gluon radiation spectrum in the medium~\cite{Blaizot:2015lma}. Much larger lattices are necessary to further investigate the effect, which is beyond our current scope of study.\footnote{In a closely-related classical study using as large as $N_\perp=16$ and $N_\eta=4$, it is found that the gluon probability depends on the medium strength and is enhanced at the presence of the medium \cite{Li:2023jeh}.} 

In Fig.~\ref{fig:medium_evo_by_z} we compute the splitting function in the medium and compare it to the estimated vacuum splitting function $P_{q\to qg}$; see also Fig.~\ref{fig:gluon_prob_vacuum_g10_zg}.  For the simulated time duration, we observe that the in-medium data points are compatible with the vacuum splitting kernel. Of course, the possible existence of deviations is shadowed by the small number of data points and the unitarity constraint. We observe larger deviations with respect to the quark probability, with the medium leading to a suppression of the single quark sector. Due to probability conservation, this indicates an excess in the gluon production due to the propagation in the medium. This is in agreement with previous studies, where it is observed that the medium can promote the production of a large amount of radiation at larger. However, understanding the origin of this radiation requires making a differential measurement in transverse space, which requires a larger lattice to resolve the distribution.

%%%%%%%%%%%%%%%%%%
\subsection{Quark entropy}
\label{sec:res_entropy}
%%%%%%%%%%%%%%%%%
In our simulations we can directly access the final jet state, and therefore easily compute the associated entropy. Here we are particularly interested in computing the entropy of the reduced density matrix of the single quark. This provides a simple and straightforward way to understand the role played by radiative corrections~\cite{Ghiglieri:2022gyv,Caucal:2021lgf,Liou:2013qya,Blaizot:2019muz} in in-medium jet evolution.\footnote{Note that such a study is more complex when performed at the level of the momentum broadening distribution discussed above.}

In what follows, we study the von Neuman (vN) entropy $S_{\rm vN}$ of the quark component of the jet. At leading order in the strong coupling, and using the single momentum mode initial condition we consider in this work, it can be shown that the quark entropy is related to the classical phase space explored by the state~\cite{Barata:2023uoi}. Since for a single particle $\langle p_\perp^2(t)\rangle \propto \hat q t$, one has that entropy should grow logarithmically with time. However, once radiation is included, the growth rate should increase.

The vN entropy of the quark component is defined as
\begin{align}
    S_{\rm vN} (x^+) &= - \mathrm{Tr}[\rho(x^+) \log_2 \rho(x^+)] \,.
\end{align}
The reduced quark density matrix is understood as being averaged over medium configurations, i.e., $\llangle{\rho (x^+)}\rrangle$. 
This averaging removes the medium’s degrees of freedom.
For a jet state in the $\ket{q}$ space, the single-event density matrix is given by
\begin{align}
\rho (x^+)= \ket{\psi(x^+)}\bra{\psi(x^+)}\;,
\end{align}
in which $\ket{\psi(x^+)}$ is the state vector.
For a jet state in the $\ket{q}+\ket{qg}$ space, we trace over the gluon degrees of freedom, 
\begin{align}
\rho(x^+)= \Tr_g(\ket{\psi(x^+)}\bra{\psi(x^+)})\;.
\end{align}
In practice, at the level of the circuit introduced in Sec.~\ref{sec:basis_encoding}, this can be achieved by performing projective measurements over the $\ket{\zeta}\otimes\ket{g}$ registers, or equivalently taking the partial trace of the full density matrix with {\tt partial\_trace} in {\tt Qiskit}~\cite{Qiskit}.

\begin{figure}[t]
    \centering
    \subfigure[\label{fig:entropy_q} $S_{\rm vN}$ in the Fock $\ket{q}$ with medium $g^2\mu=0.1$ $\mathrm{GeV}^{3/2}$
    ]{
    \raisebox{0\height}{ \includegraphics[width=0.5\textwidth]{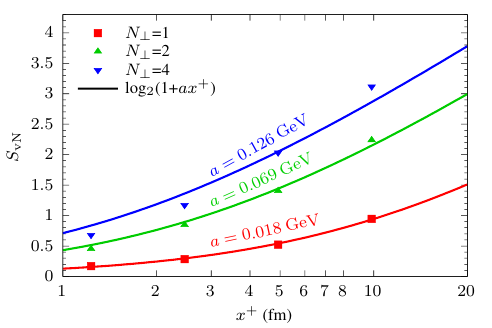}}
    }
    \quad
    \subfigure[ \label{fig:entropy_q_with_lattice} Entropy parameter $a$ as a function of $(g^2\mu)^2$
    ]{
    \includegraphics[width=0.5\textwidth]{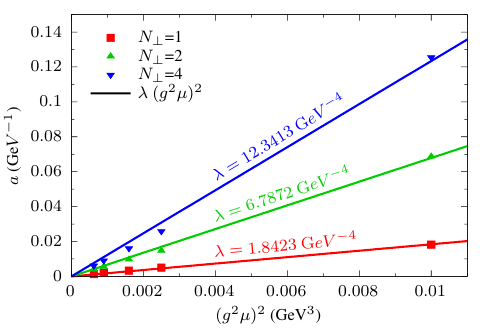}}
    \caption{Entropy growth of the quark state in the $\ket{q}$ Fock space with initial energy $p^+=1$ GeV.
    (a) Time evolution of the von Neuman entropy $S_{\rm vN}$ at various lattice sizes. (b) Entropy parameter $a$ as a function of $(g^2\mu)^2$ at various lattice sizes.  
    }
    \label{fig:entropy_main_q}
\end{figure}

We study the entropy of the jet state formulated in both the $\ket{q}$ and the $\ket{q}+\ket{qg}$ spaces.
In the single parton case, the entropy is expected to behave as $\log_2(1+a x^+)$ according to Ref.~\cite{Barata:2023uoi}, in which $a$ is a parameter related to the average transverse momentum square acquired due to the interactions with the medium. In Fig.~\ref{fig:entropy_q}, we present the simulation results for increasing lattice sizes of $N_\perp=1,2,4$ at fixed medium strength $g^2\mu = 0.1\, {\rm GeV}^{3/2}$ and finite energy $p^+=1$ GeV. We can see a larger entropy growth with the lattice size, which is expected as the phase space becomes larger. We also notice the apparent logarithmic growth for the different parameter sets used as a function of $x^+$. To further examine the dependence, we fit the data points to the expected functional form above, using $a$ as a free fitting parameter for the different $N_\perp$. 
Since $a$ is related to the average momentum transfer experienced by the quark, one expects it to grow linearly with $(g^2\mu)^2$, i.e., $\hat q$. In Fig.~\ref{fig:entropy_q_with_lattice} we show the evolution of the fitting parameter as a function of $(g^2\mu)^2$ for the different $N_\perp$ values considered. Indeed, we observe that the evolution for each lattice is reasonably described by linear regression.

When the gluon is included, the entropy can not only grow due to momentum diffusion but also as a consequence of the recoil experienced by the quark due to the gluon production. As a result, one should expect a larger growth of the associated von-Neumann entropy. 
We consider two mechanisms in describing such growth.

The first possible mechanism is that including the gluon production can lead to a larger effective $\hat q $, and therefore a larger value for the fitting parameter $a$. 
We test this hypothesis by fitting the same functional form $\log_2(1+a x^+)$ to the quark entropy computed for the two-parton scenario, and found that such a fit can not properly describe the results obtained from the simulation.
This suggests that for the quark entropy, the effect of having gluon radiation can not be reduced to having a larger effective value for $\hat q$. 

\begin{figure}[t]
    \centering
    \includegraphics[width=0.5\textwidth]{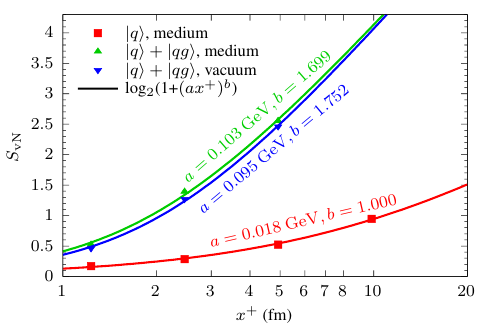}
    \caption{Entropy of the quark component $S_{\rm vN}$ in the $\ket{q}+\ket{qg}$ Fock space, in vacuum and in medium, with initial energy $p^+=1$ GeV. 
    The result in the $\ket{q}$ Fock space is plotted for comparison. Medium strength is fixed as $g^2\mu =0.1\, \mathrm{GeV}^{3/2}$. 
    The data point at $x^+=10\,\mathrm{fm}$ is affected by the lattice boundary and therefore not shown here.
    }
    \label{fig:entropy_main_qg}
\end{figure}

We then tend to the second possibility that the production of radiation can lead to an accelerated entropy growth reflected in the (anomalous) time exponent. To examine, we consider the functional form $\log_2[1+(a {x^+})^{b}]$, with $b>1$ to fit the simulation results for comparison. 
We present the data points and the fits in Fig.~\ref{fig:entropy_main_qg}, including both the case where the jet evolves in the medium and in the vacuum.\footnote{Notice that for the single particle in vacuum, $S_{\rm vN}=0$.} We observe that this functional form can properly capture data points. The fitting parameters for the medium and vacuum points are compatible and show that the gluon production mechanism is dominant over the medium effects, for the parameter sets used. 
However, we can not extract further dependencies of the fitting parameters due to numerical limitations. As a result, it is not possible at the moment to further pin down the physical meaning of the fitting parameter values obtained. We leave further research on this topic for future work.

\section{Conclusion and outlook }
\label{sec:conclusion}
%%%%%%%%%%%%%%%%%
In this work, we have implemented a digital quantum circuit to quantum simulate the evolution of a QCD jet.
We implement the light-front Hamiltonian formalism and perform the real-time quantum simulation in the $\ket{q}$ and the $\ket{q}+\ket{qg}$ Fock spaces. We have studied the total momentum broadening of the jet, the gluon production, and the von Neumann entropy associated to the quark state. 

We find sizable eikonal effects by comparing the total momentum broadening of the jet in the $\ket{q}+\ket{qg}$ space at finite energy, to that in the $\ket{q}$ space at the eikonal limit of $p^+=\infty$. 
The underlying physics is that the inclusion of gluon radiation, which depends on $p^+$, enlarges the phase space, and the medium interaction also interferes with this process.
Furthermore, when studying the energy distribution of the gluon inside the jet, we recover the reduced vacuum splitting function. When the medium is included, we can not observe significant modifications to the vacuum kernel, but note that there is a larger amount of gluon radiation being produced. 
We leave the interesting study on the closely related QCD LPM effect for future work when simulations on larger lattices are feasible.

Finally, we compute the entropy associated with the quark state, in both the $\ket{q}$ and the $\ket{q}+\ket{qg}$ Fock spaces. In the former case, we recover the classical result for momentum diffusion, in which the entropy grows logarithmically in time. In the latter case, the entropy growth accelerates significantly, mainly due to the production of gluon radiation, irrespective of the medium being present.

Extensions of the algorithm presented here to include higher Fock sectors, such as $\ket{qgg} $, are underway. We note that including this sector would allow to perform numerical calculations beyond known analytical results, see e.g.~\cite{Arnold:2023qwi,Arnold:2015qya}. Another interesting avenue to be explored regards to the transition of the prepared final state partonic jet into a hadronic state, such as a pion state obtained on the circuit~\cite{Qian:2021jxp}. 

\section*{Acknowledgments}
We are grateful to Tuomas Lappi, James P. Vary, Xin-Nian Wang, Bin Wu, and Xingbo Zhao for their helpful and valuable discussions. We acknowledge the use of IBM Quantum services for this work. The views expressed are those of the authors and do not reflect the official policy or position of IBM or the IBM Quantum team. 
JB is supported by the U.S. Department of Energy, Office of Science, National Quantum Information Science Research Centers under the “Co-design Center for Quantum Advantage” award and by the U.S. Department of Energy, Office of Science, Office of Nuclear Physics, under contract No. DE-SC0012704.
XD, ML, WQ, and CS are supported by Xunta de Galicia (Centro singular de investigacion de Galicia accreditation 2019-2022), European Union ERDF, the “Maria de Maeztu” Units of Excellence program under project CEX2020-001035-M, the Spanish Research State Agency under project PID2020-119632GB-I00, and European Research Council under project ERC-2018-ADG-835105 YoctoLHC.
WQ is also supported by funding from the European Union's Marie Skłodowska-Curie Actions Postdoctoral Fellowships 2022 (HORIZON-MSCA-2022-PF-01) under Grant Agreement No. 101109293.

%%%%%%%%%%%%%%%%%%
\appendix
%%%%%%%%%%%%%%%%%%

\section{Computation of the background field}~\label{app:field}

In this appendix, we present the computation of the background field in both the transverse momentum and the transverse position spaces on the lattice, as we have implemented in the simulations. We follow the approach in Refs.~\cite{Li:2020uhl, Li:2021zaw}.

In the discrete basis space, the correlation relation of the color charge density in Eq.~\eqref{eq:MV_color_charge} takes the form,
\begin{multline}\label{eq:chgcor_dis}
  \braket{\rho_a(n^x,n^y,n_\tau)\rho_b({n'}^x,{n'}^y,n_\tau')}\\
 =g^2 \mu^2\delta_{ab}\frac{\delta_{n^x,{n'}^x}\delta_{n^y,{n'}^y}}{a_\perp^2}\frac{\delta_{n_\tau,n_\tau'}}{\tau}\;.
\end{multline}
The sources generating the medium are stochastic random variables with a Gaussian distribution on each site, with the transverse indices $n^x,n^y=-N_\perp,-N_\perp+1,\ldots, N_\perp-1$, and the layer indices $n_\tau=1,2,\ldots, N_\eta$. 
The charge density in the momentum space is obtained by the Fourier transform
$\tilde \rho_a(k^x, k^y, n_\tau)
=\sum_{\bar n_x, \bar n_y=-N_\perp}^{N_\perp-1} 
\rho_a(\bar n^x,\bar n^y, n_\tau)
e^{i(\bar n_x k_x + \bar n_y  k_y)\pi/N_\perp} $.

The field equation as given in Eq.~\eqref{eq:MV} is straightforward to solve in the momentum space, then the field in the coordinate space can be obtained by an inverse Fourier transform,
 \begin{align}\label{eq:CGCA}
  \begin{split}
\tilde {\mathcal A}^-_a (k^x, k^y, x^+)
=&
\frac{\tilde \rho_a(k^x, k^y, n_\tau(x^+)) }{m_g^2 a_\perp^2/\pi^2/N_\perp^2+k_x^2+k_y^2} 
\;,\\
\mathcal{A}^-_a (n^x,n^y, x^+)
=&
\frac{\sum_{ k_x, k_y=-N_\perp}^{N_\perp-1}}{{(2N_\perp)}^2}
\tilde {\mathcal A}^-_a (k^x, k^y, x^+)
\\
&
 e^{-i(n_x k_x + n_y k_y)\pi/N_\perp}  
\;.
\end{split}
\end{align}
We write $n_\tau(x^+)$ to indicate that the layer indices $n_\tau$ can be determined by the position of $x^+$ in the entire duration of $[0, L_\eta]$.
For each layer, $\rho$ is sampled independently, so the resulting $\mathcal{A}^-$ is uncorrelated across layers. 

\section{Conventions on SU(2) color structure}\label{app:SU2_algebra}

The SU(2) algebra is
\begin{align}
    [J_j, J_k] = i\epsilon_{jkl} J_l
    \;,
\end{align}
where $\epsilon_{jkl}=1(-1)$ for even (odd) permutations of $\{j,k,l\}=\{1,2,3\}$, otherwise 0. 

The generators in the fundamental representation, denoted by $T$, are
\begin{align}
    T_1=\frac{\sigma^X}{2},\;
    T_2=\frac{\sigma^Y}{2},\;
    T_3=\frac{\sigma^Z}{2},
\end{align}
where the Pauli matrices are defined as
\begin{align}
    \sigma^X = \begin{pmatrix}
    0&1\\1&0
    \end{pmatrix},\;
    \sigma^Y = \begin{pmatrix}
    0&-i\\i&0
    \end{pmatrix},\;
    \sigma^Z = \begin{pmatrix}
    1&0\\0&-1
    \end{pmatrix}.
\end{align}

The generators in the adjoint representation, the structure constant $\epsilon$, are
\begin{align}
\begin{split}
    &\epsilon_{1bc}=
    \begin{pmatrix}
    0 & 0 & 0\\
    0 & 0 & 1\\
    0 & -1 & 0
    \end{pmatrix},
    \epsilon_{2bc}=
    \begin{pmatrix}
    0 & 0 & -1\\
    0 & 0 & 0\\
    1 & 0 & 0
    \end{pmatrix},\\
    &
    \epsilon_{3bc}= 
    \begin{pmatrix}
    0 &1  & 0\\
    -1 & 0 & 0\\
    0 &0  & 0
    \end{pmatrix}\;.
    \end{split}
\end{align}
These matrices can be efficiently represented by Pauli strings when encoded to a 2-qubit register,
\begin{align}
\begin{split}
    \epsilon_{1bc}&=
    -0.5i (\sigma^X \sigma^Y - \sigma^Y \sigma^X),\\
    \epsilon_{2bc}&=
    -0.5i (\sigma^Y \sigma^I + \sigma^Y \sigma^Z), \\
    \epsilon_{3bc}&=
    0.5i (\sigma^I \sigma^Y + \sigma^Z \sigma^Y)\;.
\end{split}
\end{align}

\section{Evaluation of the Pauli terms}\label{app:Pauli_strings}

In this work, we will simply decompose the Hamiltonian matrix into a sum of Pauli operators and then evolve them in time. The efficient acquisition of the Pauli strings is particularly important. For this purpose, we look closely at the two implementations.

\begin{enumerate}
    \item Orthogonal Matrix Projection (omp) 
    
    The industry standard approach is to use the orthogonal projection of the Hamiltonian matrix onto each possible Pauli matrix. For any Hermitian matrix $H$ of size $2^n$-by-$2^n$, the general decomposition~\cite{nielsen_chuang_2010} can be expressed as
    \begin{align}\label{eq:pauli_trace_decomp}
        H &= \sum_x \omega_x \big(\sigma_{x_n}\otimes \cdots \otimes \sigma_{x_2}\otimes \sigma_{x_1}\big) \\
        &\equiv \sum_x \omega_x P(x),
    \end{align}
    where $x=x_n\cdots x_2x_1=\{0,1,2,3\}^n$ and $\sigma_{0,1,2,3}=\{I, \sigma^X, \sigma^Y, \sigma^Z\}$ is a collection of the Pauli matrices. All the non-zero weights $\omega_x= \frac{1}{2^n}\mathrm{Tr}\big[P(x) H\big]$ of the associated Pauli string can be obtained. Despite recent efforts~\cite{Romero:2023mob} taking advantage of the properties of the Pauli matrices as well as multi-core parallelization, this method is generally very inefficient in dealing with very sparse matrices that we have in our mixed-space simulation. 

    \item Sparse Matrix Projection (smp)
    
    Alternatively, one can directly evaluate the Pauli strings for each non-zero matrix element\footnote{The sparse matrix representation of the matrix elements for each term of the Hamiltonian is easily prepared beforehand.} using the Boolean function operator $f_{\mathcal{B}}(i, j)$~\cite{hadfield:2021}. For each non-zero matrix element $h_{ij}$ with $i,j=1,...,2^n$, its related set of the Pauli strings $P(h_{ij})$ are
    \begin{align}\label{eq:pauli_boolean}
       h_{ij} f_{\mathcal{B}}(i, j) \equiv h_{i,j} \big(\otimes^{n}_{k=1}f_{\mathcal{B}}(i_k, j_k)\big),
    \end{align}
    where $i_k,j_k=0,1$ is the k-th digit of the binary representation
    of matrix index $i, j$ and on a single bit $f_{\mathcal{B}}$ is 
    \begin{align}\label{eq:pauli_boolean_1bit}
    \begin{split}
        &f_{\mathcal{B}}(0,0) = \frac{I+\sigma^Z}{2} \;, \\
        &f_{\mathcal{B}}(1,1) = \frac{I-\sigma^Z}{2}\;,\\ 
        &f_{\mathcal{B}}(1,0) = \frac{\sigma^X-i\sigma^Y}{2} \;,\\
        &f_{\mathcal{B}}(0,1) =\frac{\sigma^X+i\sigma^Y}{2}\;.
    \end{split}
    \end{align}
    Simply put, we evaluate each set of nonzero Pauli strings as a tensor product of transition operators between $\{\ket{0}, \ket{1}\}$. Then, the final Hamiltonian at the end is a reduced sum of these Pauli strings. This sparse matrix projection strategy works exceptionally well for the mixed-space unitary evolution since the sparsity of these matrices is well below 1\%. 
\end{enumerate}
In Table.~\ref{tab:pauli}, we present and compare the resource cost of evaluating Pauli terms with both approaches for the two evolution methods in Sect.~\ref{sec:gate_encoding} in the paper. We can see that the omp method does not scale with increasing problem size. Within the smp methods, there is a considerable advantage of the mixed-space representation over the momentum-space representation, mostly due to the sparsity and the diagonality of the $V_\mathcal{A}$ terms. We notice that the recombination of all the Pauli terms in the smp method also becomes costly as the problem size increases.

\begin{table*}
  \centering
  \caption{\label{tab:pauli}Resource cost of evaluating various Pauli terms with both the orthogonal matrix projection (omp) and the sparse matrix projection (smp) methods. We compare the computational time costs (in seconds) needed for the momentum-space ($t^P$) and the mixed-space strategies ($t^M$). The respective sparsities ($\mathcal{S}$) of the medium interaction matrix $V_\mathcal{A}$ in each method are also provided. Numerical benchmark results are performed on the same Ubuntu 22.04.2 LTS machine using 1 CPU core with 32.0 GB memory and an Intel i9 processor of 3.50 GHz. Ideally, both the omp and smp methods can be parallelized using multi-core CPUs.  
  } 
    \begin{tabular}
    % {c c l l l l l l l  l}
    {  c@{\hskip 0.1in} | c@{\hskip 0.1in} | c@{\hskip 0.1in} ||  c@{\hskip 0.2in} | c@{\hskip 0.1in}  c@{\hskip 0.2in} | c@{\hskip 0.1in} c@{\hskip 0.1in} c@{\hskip 0.1in} c@{\hskip 0.1in} }
    \toprule
\toprule
\hline
\hline
    $N_\perp$
    & $\lceil K \rceil$
    & $n_Q$
    & $t^{P}_\mathrm{omp} (s)$ 
    & $\mathcal{S}^{P}_{V_{\mathcal{A}}}$
    & $t^{P}_\mathrm{smp, H_\mathrm{tot}} (s)$ 
    & $t^{M}_\mathrm{smp, K+V_{qg}} (s)$ 
    & $t^{M}_\mathrm{smp, V_\mathcal{A}} (s)$ 
    & $\mathcal{S}^{M}_{V_\mathcal{A}}$
    & $t^{M}_\mathrm{smp, H_\mathrm{tot}} (s)$ 
    \\
     \colrule
    1 & 2 & 8 &14.86   &2.44\% &1.69  &0.29 &0.38 &0.61\%  &0.67\\
     \hline 
    1 & 4 & 9 &82.30   &1.78\% &11.83 &1.23 &1.60 &0.44\%  &2.83\\   
    \hline 
    1 & 8 & 10 &501.20 &1.03\%  &149.71 &5.27 &7.29 &0.26\% &12.57\\
    \hline\hline 
    2 & 2 & 12 &-\footnotemark[1] &0.59\% &20545.68\footnotemark[2] &111.67 &174.21 &0.04\% &285.88\\
    \hline
    2 & 4 & 13 &-\footnotemark[1] &0.44\% &137278.40\footnotemark[2] &1921.39 &3070.85 &0.03\% &4992.23\\
     \botrule
    \end{tabular}\\
    \raggedright
    \footnotemark[1]{Using the omp method at $N_\perp=2$ takes an extremely long time, so they are not presented.}\\
    \footnotemark[2]{We used a fraction of the Pauli terms to estimate the total time because the full calculations are very expensive. Their true time costs are expected to be much larger.}
\end{table*}
 
%%%%%%%%%%%%%%%%%%
\section{Convergence on $N_\eta$ and $N_{dx}$}\label{app:NtNeta_converge}
%%%%%%%%%%%%%%%%%%

The MV model that describes the background field is formulated in the continuous limit of $N_\eta \to\infty$, thus also the corresponding analytical expectation on $\braket{p_\perp^2}$. In numerical simulations, one takes finite values of $N_\eta$, which can lead to layer effects; see more discussions in Refs.~\cite{Barata:2022wim, Li:2023jeh}. 
Here, we use a sufficiently large value of $N_\eta$ such that in the $p^+=\infty$ limit,  the quantum simulation result agrees with analytical expectation in the $N_\eta \to\infty$ limit.

\begin{figure*}[t]
    \centering
    \subfigure[\label{fig:convergence_Neta} $N_\eta$ convergence at $p^+=\infty$.
    ]{
    \raisebox{0\height}{ \includegraphics[width=0.45\textwidth]{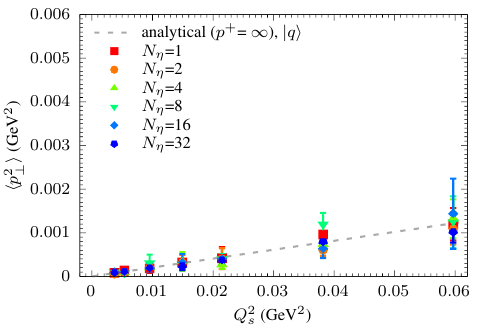}
    }}
    \quad
    \subfigure[ \label{fig:convergence_Ndx} $N_{t}$ convergence at $p^+=1\, \mathrm{GeV}, N_\eta=4,\, g^2\mu=0.1 \mathrm{GeV}^{3/2}$.
    ]{
\includegraphics[width=0.45\textwidth]{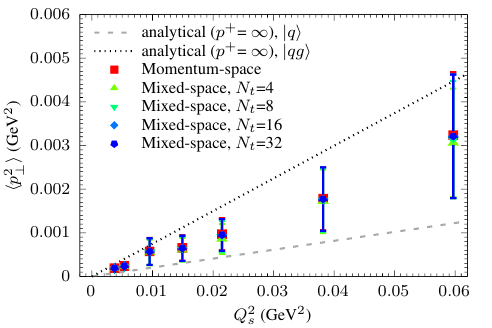}}
    \caption{Convergence studies of the number of layers ($N_\eta$) used in the MV model and time steps ($N_t$) in the trotterizations.
    }
    \label{fig:p2_convergence}
\end{figure*}

We use $N_\eta=4$ for the results presented in the main body of this paper. Here, we show the results of $\braket{p_\perp^2}$ at various $N_\eta$ in the eikonal limit of $p^+=\infty$ limit
In Fig.~\ref{fig:convergence_Neta}, we show the transverse momenta at increasing $Q_s^2$ and notice that our results agree with analytical results even at $N_\eta=1$. 

We also examine the convergence of the trotterization step size $N_t$. In particular, we compare the in-medium momentum broadening using the two simulation strategies (momentum-space vs mixed-space) in Fig.~\ref{fig:convergence_Ndx} for a finite initial quark energy, i.e., $p^+=1\, \mathrm{GeV}$. We can see that any $N_t$ could give reasonable results.
At the value of $N_t$ used in this work $N_t=16$, the simulation result is within $1\%$ of the expected value. 

%%%%%%%%%%%%%%%%%%
\section*{}
\bibliographystyle{apsrev4-1}
\bibliography{mybib.bib}

\begin{thebibliography}{56}%
\makeatletter
\providecommand \@ifxundefined [1]{%
 \@ifx{#1\undefined}
}%
\providecommand \@ifnum [1]{%
 \ifnum #1\expandafter \@firstoftwo
 \else \expandafter \@secondoftwo
 \fi
}%
\providecommand \@ifx [1]{%
 \ifx #1\expandafter \@firstoftwo
 \else \expandafter \@secondoftwo
 \fi
}%
\providecommand \natexlab [1]{#1}%
\providecommand \enquote  [1]{``#1''}%
\providecommand \bibnamefont  [1]{#1}%
\providecommand \bibfnamefont [1]{#1}%
\providecommand \citenamefont [1]{#1}%
\providecommand \href@noop [0]{\@secondoftwo}%
\providecommand \href [0]{\begingroup \@sanitize@url \@href}%
\providecommand \@href[1]{\@@startlink{#1}\@@href}%
\providecommand \@@href[1]{\endgroup#1\@@endlink}%
\providecommand \@sanitize@url [0]{\catcode `\\12\catcode `\$12\catcode
  `\&12\catcode `\#12\catcode `\^12\catcode `\_12\catcode `\%12\relax}%
\providecommand \@@startlink[1]{}%
\providecommand \@@endlink[0]{}%
\providecommand \url  [0]{\begingroup\@sanitize@url \@url }%
\providecommand \@url [1]{\endgroup\@href {#1}{\urlprefix }}%
\providecommand \urlprefix  [0]{URL }%
\providecommand \Eprint [0]{\href }%
\providecommand \doibase [0]{http://dx.doi.org/}%
\providecommand \selectlanguage [0]{\@gobble}%
\providecommand \bibinfo  [0]{\@secondoftwo}%
\providecommand \bibfield  [0]{\@secondoftwo}%
\providecommand \translation [1]{[#1]}%
\providecommand \BibitemOpen [0]{}%
\providecommand \bibitemStop [0]{}%
\providecommand \bibitemNoStop [0]{.\EOS\space}%
\providecommand \EOS [0]{\spacefactor3000\relax}%
\providecommand \BibitemShut  [1]{\csname bibitem#1\endcsname}%
\let\auto@bib@innerbib\@empty
%</preamble>
\bibitem [{\citenamefont {Casalderrey-Solana}\ and\ \citenamefont
  {Salgado}(2007)}]{Casalderrey-Solana:2007knd}%
  \BibitemOpen
  \bibfield  {author} {\bibinfo {author} {\bibfnamefont {J.}~\bibnamefont
  {Casalderrey-Solana}}\ and\ \bibinfo {author} {\bibfnamefont {C.~A.}\
  \bibnamefont {Salgado}},\ }\href@noop {} {\bibfield  {journal} {\bibinfo
  {journal} {Acta Phys. Polon. B}\ }\textbf {\bibinfo {volume} {38}},\ \bibinfo
  {pages} {3731} (\bibinfo {year} {2007})},\ \Eprint
  {http://arxiv.org/abs/0712.3443} {arXiv:0712.3443 [hep-ph]} \BibitemShut
  {NoStop}%
\bibitem [{\citenamefont {Majumder}\ and\ \citenamefont
  {Van~Leeuwen}(2011)}]{Majumder:2010qh}%
  \BibitemOpen
  \bibfield  {author} {\bibinfo {author} {\bibfnamefont {A.}~\bibnamefont
  {Majumder}}\ and\ \bibinfo {author} {\bibfnamefont {M.}~\bibnamefont
  {Van~Leeuwen}},\ }\href {\doibase 10.1016/j.ppnp.2010.09.001} {\bibfield
  {journal} {\bibinfo  {journal} {Prog. Part. Nucl. Phys.}\ }\textbf {\bibinfo
  {volume} {66}},\ \bibinfo {pages} {41} (\bibinfo {year} {2011})},\ \Eprint
  {http://arxiv.org/abs/1002.2206} {arXiv:1002.2206 [hep-ph]} \BibitemShut
  {NoStop}%
\bibitem [{\citenamefont {Qin}\ and\ \citenamefont {Wang}(2015)}]{Qin:2015srf}%
  \BibitemOpen
  \bibfield  {author} {\bibinfo {author} {\bibfnamefont {G.-Y.}\ \bibnamefont
  {Qin}}\ and\ \bibinfo {author} {\bibfnamefont {X.-N.}\ \bibnamefont {Wang}},\
  }\href {\doibase 10.1142/S0218301315300143} {\bibfield  {journal} {\bibinfo
  {journal} {Int. J. Mod. Phys. E}\ }\textbf {\bibinfo {volume} {24}},\
  \bibinfo {pages} {1530014} (\bibinfo {year} {2015})},\ \Eprint
  {http://arxiv.org/abs/1511.00790} {arXiv:1511.00790 [hep-ph]} \BibitemShut
  {NoStop}%
\bibitem [{\citenamefont {Apolin\'ario}\ \emph {et~al.}(2022)\citenamefont
  {Apolin\'ario}, \citenamefont {Lee},\ and\ \citenamefont
  {Winn}}]{Apolinario:2022vzg}%
  \BibitemOpen
  \bibfield  {author} {\bibinfo {author} {\bibfnamefont {L.}~\bibnamefont
  {Apolin\'ario}}, \bibinfo {author} {\bibfnamefont {Y.-J.}\ \bibnamefont
  {Lee}}, \ and\ \bibinfo {author} {\bibfnamefont {M.}~\bibnamefont {Winn}},\
  }\href {\doibase 10.1016/j.ppnp.2022.103990} {\bibfield  {journal} {\bibinfo
  {journal} {Prog. Part. Nucl. Phys.}\ }\textbf {\bibinfo {volume} {127}},\
  \bibinfo {pages} {103990} (\bibinfo {year} {2022})},\ \Eprint
  {http://arxiv.org/abs/2203.16352} {arXiv:2203.16352 [hep-ph]} \BibitemShut
  {NoStop}%
\bibitem [{\citenamefont {Blaizot}\ and\ \citenamefont
  {Mehtar-Tani}(2015)}]{Blaizot:2015lma}%
  \BibitemOpen
  \bibfield  {author} {\bibinfo {author} {\bibfnamefont {J.-P.}\ \bibnamefont
  {Blaizot}}\ and\ \bibinfo {author} {\bibfnamefont {Y.}~\bibnamefont
  {Mehtar-Tani}},\ }\href {\doibase 10.1142/S021830131530012X} {\bibfield
  {journal} {\bibinfo  {journal} {Int. J. Mod. Phys. E}\ }\textbf {\bibinfo
  {volume} {24}},\ \bibinfo {pages} {1530012} (\bibinfo {year} {2015})},\
  \Eprint {http://arxiv.org/abs/1503.05958} {arXiv:1503.05958 [hep-ph]}
  \BibitemShut {NoStop}%
\bibitem [{\citenamefont {Gyulassy}\ \emph {et~al.}(2001)\citenamefont
  {Gyulassy}, \citenamefont {Levai},\ and\ \citenamefont
  {Vitev}}]{Gyulassy:2000er}%
  \BibitemOpen
  \bibfield  {author} {\bibinfo {author} {\bibfnamefont {M.}~\bibnamefont
  {Gyulassy}}, \bibinfo {author} {\bibfnamefont {P.}~\bibnamefont {Levai}}, \
  and\ \bibinfo {author} {\bibfnamefont {I.}~\bibnamefont {Vitev}},\ }\href
  {\doibase 10.1016/S0550-3213(00)00652-0} {\bibfield  {journal} {\bibinfo
  {journal} {Nucl. Phys. B}\ }\textbf {\bibinfo {volume} {594}},\ \bibinfo
  {pages} {371} (\bibinfo {year} {2001})},\ \Eprint
  {http://arxiv.org/abs/nucl-th/0006010} {arXiv:nucl-th/0006010} \BibitemShut
  {NoStop}%
\bibitem [{\citenamefont {Zakharov}(1996)}]{Zakharov:1996fv}%
  \BibitemOpen
  \bibfield  {author} {\bibinfo {author} {\bibfnamefont {B.~G.}\ \bibnamefont
  {Zakharov}},\ }\href {\doibase 10.1134/1.567126} {\bibfield  {journal}
  {\bibinfo  {journal} {JETP Lett.}\ }\textbf {\bibinfo {volume} {63}},\
  \bibinfo {pages} {952} (\bibinfo {year} {1996})},\ \Eprint
  {http://arxiv.org/abs/hep-ph/9607440} {arXiv:hep-ph/9607440} \BibitemShut
  {NoStop}%
\bibitem [{\citenamefont {Baier}\ \emph {et~al.}(1997)\citenamefont {Baier},
  \citenamefont {Dokshitzer}, \citenamefont {Mueller}, \citenamefont {Peigne},\
  and\ \citenamefont {Schiff}}]{Baier:1996sk}%
  \BibitemOpen
  \bibfield  {author} {\bibinfo {author} {\bibfnamefont {R.}~\bibnamefont
  {Baier}}, \bibinfo {author} {\bibfnamefont {Y.~L.}\ \bibnamefont
  {Dokshitzer}}, \bibinfo {author} {\bibfnamefont {A.~H.}\ \bibnamefont
  {Mueller}}, \bibinfo {author} {\bibfnamefont {S.}~\bibnamefont {Peigne}}, \
  and\ \bibinfo {author} {\bibfnamefont {D.}~\bibnamefont {Schiff}},\ }\href
  {\doibase 10.1016/S0550-3213(96)00581-0} {\bibfield  {journal} {\bibinfo
  {journal} {Nucl. Phys. B}\ }\textbf {\bibinfo {volume} {484}},\ \bibinfo
  {pages} {265} (\bibinfo {year} {1997})},\ \Eprint
  {http://arxiv.org/abs/hep-ph/9608322} {arXiv:hep-ph/9608322} \BibitemShut
  {NoStop}%
\bibitem [{\citenamefont {Arnold}\ and\ \citenamefont
  {Iqbal}(2015)}]{Arnold:2015qya}%
  \BibitemOpen
  \bibfield  {author} {\bibinfo {author} {\bibfnamefont {P.}~\bibnamefont
  {Arnold}}\ and\ \bibinfo {author} {\bibfnamefont {S.}~\bibnamefont {Iqbal}},\
  }\href {\doibase 10.1007/JHEP09(2016)072} {\bibfield  {journal} {\bibinfo
  {journal} {JHEP}\ }\textbf {\bibinfo {volume} {04}},\ \bibinfo {pages} {070}
  (\bibinfo {year} {2015})},\ \bibinfo {note} {[Erratum: JHEP 09, 072
  (2016)]},\ \Eprint {http://arxiv.org/abs/1501.04964} {arXiv:1501.04964
  [hep-ph]} \BibitemShut {NoStop}%
\bibitem [{\citenamefont {Fickinger}\ \emph {et~al.}(2013)\citenamefont
  {Fickinger}, \citenamefont {Ovanesyan},\ and\ \citenamefont
  {Vitev}}]{Fickinger:2013xwa}%
  \BibitemOpen
  \bibfield  {author} {\bibinfo {author} {\bibfnamefont {M.}~\bibnamefont
  {Fickinger}}, \bibinfo {author} {\bibfnamefont {G.}~\bibnamefont
  {Ovanesyan}}, \ and\ \bibinfo {author} {\bibfnamefont {I.}~\bibnamefont
  {Vitev}},\ }\href {\doibase 10.1007/JHEP07(2013)059} {\bibfield  {journal}
  {\bibinfo  {journal} {JHEP}\ }\textbf {\bibinfo {volume} {07}},\ \bibinfo
  {pages} {059} (\bibinfo {year} {2013})},\ \Eprint
  {http://arxiv.org/abs/1304.3497} {arXiv:1304.3497 [hep-ph]} \BibitemShut
  {NoStop}%
\bibitem [{\citenamefont {Barata}\ and\ \citenamefont
  {Salgado}(2021)}]{Barata:2021yri}%
  \BibitemOpen
  \bibfield  {author} {\bibinfo {author} {\bibfnamefont {J.}~\bibnamefont
  {Barata}}\ and\ \bibinfo {author} {\bibfnamefont {C.~A.}\ \bibnamefont
  {Salgado}},\ }\href {\doibase 10.1140/epjc/s10052-021-09674-9} {\bibfield
  {journal} {\bibinfo  {journal} {Eur. Phys. J. C}\ }\textbf {\bibinfo {volume}
  {81}},\ \bibinfo {pages} {862} (\bibinfo {year} {2021})},\ \Eprint
  {http://arxiv.org/abs/2104.04661} {arXiv:2104.04661 [hep-ph]} \BibitemShut
  {NoStop}%
\bibitem [{\citenamefont {De~Jong}\ \emph {et~al.}(2021)\citenamefont
  {De~Jong}, \citenamefont {Metcalf}, \citenamefont {Mulligan}, \citenamefont
  {{P\l osko\'n}}, \citenamefont {Ringer},\ and\ \citenamefont
  {Yao}}]{DeJong:2020riy}%
  \BibitemOpen
  \bibfield  {author} {\bibinfo {author} {\bibfnamefont {W.~A.}\ \bibnamefont
  {De~Jong}}, \bibinfo {author} {\bibfnamefont {M.}~\bibnamefont {Metcalf}},
  \bibinfo {author} {\bibfnamefont {J.}~\bibnamefont {Mulligan}}, \bibinfo
  {author} {\bibfnamefont {M.}~\bibnamefont {{P\l osko\'n}}}, \bibinfo {author}
  {\bibfnamefont {F.}~\bibnamefont {Ringer}}, \ and\ \bibinfo {author}
  {\bibfnamefont {X.}~\bibnamefont {Yao}},\ }\href {\doibase
  10.1103/PhysRevD.104.L051501} {\bibfield  {journal} {\bibinfo  {journal}
  {Phys. Rev. D}\ }\textbf {\bibinfo {volume} {104}},\ \bibinfo {pages}
  {051501} (\bibinfo {year} {2021})},\ \Eprint
  {http://arxiv.org/abs/2010.03571} {arXiv:2010.03571 [hep-ph]} \BibitemShut
  {NoStop}%
\bibitem [{\citenamefont {Barata}\ \emph {et~al.}(2022)\citenamefont {Barata},
  \citenamefont {Du}, \citenamefont {Li}, \citenamefont {Qian},\ and\
  \citenamefont {Salgado}}]{Barata:2022wim}%
  \BibitemOpen
  \bibfield  {author} {\bibinfo {author} {\bibfnamefont {J.~a.}\ \bibnamefont
  {Barata}}, \bibinfo {author} {\bibfnamefont {X.}~\bibnamefont {Du}}, \bibinfo
  {author} {\bibfnamefont {M.}~\bibnamefont {Li}}, \bibinfo {author}
  {\bibfnamefont {W.}~\bibnamefont {Qian}}, \ and\ \bibinfo {author}
  {\bibfnamefont {C.~A.}\ \bibnamefont {Salgado}},\ }\href {\doibase
  10.1103/PhysRevD.106.074013} {\bibfield  {journal} {\bibinfo  {journal}
  {Phys. Rev. D}\ }\textbf {\bibinfo {volume} {106}},\ \bibinfo {pages}
  {074013} (\bibinfo {year} {2022})},\ \Eprint
  {http://arxiv.org/abs/2208.06750} {arXiv:2208.06750 [hep-ph]} \BibitemShut
  {NoStop}%
\bibitem [{\citenamefont {Zhao}\ \emph {et~al.}(2013)\citenamefont {Zhao},
  \citenamefont {Ilderton}, \citenamefont {Maris},\ and\ \citenamefont
  {Vary}}]{Zhao:2013cma}%
  \BibitemOpen
  \bibfield  {author} {\bibinfo {author} {\bibfnamefont {X.}~\bibnamefont
  {Zhao}}, \bibinfo {author} {\bibfnamefont {A.}~\bibnamefont {Ilderton}},
  \bibinfo {author} {\bibfnamefont {P.}~\bibnamefont {Maris}}, \ and\ \bibinfo
  {author} {\bibfnamefont {J.~P.}\ \bibnamefont {Vary}},\ }\href {\doibase
  10.1103/PhysRevD.88.065014} {\bibfield  {journal} {\bibinfo  {journal} {Phys.
  Rev.}\ }\textbf {\bibinfo {volume} {D88}},\ \bibinfo {pages} {065014}
  (\bibinfo {year} {2013})},\ \Eprint {http://arxiv.org/abs/1303.3273}
  {arXiv:1303.3273 [nucl-th]} \BibitemShut {NoStop}%
%%CITATION = ARXIV:1303.3273;%%
\bibitem [{\citenamefont {Hu}\ \emph {et~al.}(2020)\citenamefont {Hu},
  \citenamefont {Ilderton},\ and\ \citenamefont {Zhao}}]{Hu:2019hjx}%
  \BibitemOpen
  \bibfield  {author} {\bibinfo {author} {\bibfnamefont {B.}~\bibnamefont
  {Hu}}, \bibinfo {author} {\bibfnamefont {A.}~\bibnamefont {Ilderton}}, \ and\
  \bibinfo {author} {\bibfnamefont {X.}~\bibnamefont {Zhao}},\ }\href {\doibase
  10.1103/PhysRevD.102.016017} {\bibfield  {journal} {\bibinfo  {journal}
  {Phys. Rev. D}\ }\textbf {\bibinfo {volume} {102}},\ \bibinfo {pages}
  {016017} (\bibinfo {year} {2020})},\ \Eprint
  {http://arxiv.org/abs/1911.12307} {arXiv:1911.12307 [nucl-th]} \BibitemShut
  {NoStop}%
\bibitem [{\citenamefont {Chen}\ \emph {et~al.}(2017)\citenamefont {Chen},
  \citenamefont {Zhao}, \citenamefont {Li}, \citenamefont {Tuchin},\ and\
  \citenamefont {Vary}}]{Chen:2017uuq}%
  \BibitemOpen
  \bibfield  {author} {\bibinfo {author} {\bibfnamefont {G.}~\bibnamefont
  {Chen}}, \bibinfo {author} {\bibfnamefont {X.}~\bibnamefont {Zhao}}, \bibinfo
  {author} {\bibfnamefont {Y.}~\bibnamefont {Li}}, \bibinfo {author}
  {\bibfnamefont {K.}~\bibnamefont {Tuchin}}, \ and\ \bibinfo {author}
  {\bibfnamefont {J.~P.}\ \bibnamefont {Vary}},\ }\href {\doibase
  10.1103/PhysRevD.95.096012} {\bibfield  {journal} {\bibinfo  {journal} {Phys.
  Rev.}\ }\textbf {\bibinfo {volume} {D95}},\ \bibinfo {pages} {096012}
  (\bibinfo {year} {2017})},\ \Eprint {http://arxiv.org/abs/1702.06932}
  {arXiv:1702.06932 [nucl-th]} \BibitemShut {NoStop}%
%%CITATION = ARXIV:1702.06932;%%
\bibitem [{\citenamefont {Lei}\ \emph {et~al.}(2022)\citenamefont {Lei},
  \citenamefont {Hu},\ and\ \citenamefont {Zhao}}]{Lei:2022nsk}%
  \BibitemOpen
  \bibfield  {author} {\bibinfo {author} {\bibfnamefont {Z.}~\bibnamefont
  {Lei}}, \bibinfo {author} {\bibfnamefont {B.}~\bibnamefont {Hu}}, \ and\
  \bibinfo {author} {\bibfnamefont {X.}~\bibnamefont {Zhao}},\ }\href@noop {}
  {\  (\bibinfo {year} {2022})},\ \Eprint {http://arxiv.org/abs/2201.01746}
  {arXiv:2201.01746 [hep-ph]} \BibitemShut {NoStop}%
\bibitem [{\citenamefont {Li}\ \emph {et~al.}(2020)\citenamefont {Li},
  \citenamefont {Zhao}, \citenamefont {Maris}, \citenamefont {Chen},
  \citenamefont {Li}, \citenamefont {Tuchin},\ and\ \citenamefont
  {Vary}}]{Li:2020uhl}%
  \BibitemOpen
  \bibfield  {author} {\bibinfo {author} {\bibfnamefont {M.}~\bibnamefont
  {Li}}, \bibinfo {author} {\bibfnamefont {X.}~\bibnamefont {Zhao}}, \bibinfo
  {author} {\bibfnamefont {P.}~\bibnamefont {Maris}}, \bibinfo {author}
  {\bibfnamefont {G.}~\bibnamefont {Chen}}, \bibinfo {author} {\bibfnamefont
  {Y.}~\bibnamefont {Li}}, \bibinfo {author} {\bibfnamefont {K.}~\bibnamefont
  {Tuchin}}, \ and\ \bibinfo {author} {\bibfnamefont {J.~P.}\ \bibnamefont
  {Vary}},\ }\href {\doibase 10.1103/PhysRevD.101.076016} {\bibfield  {journal}
  {\bibinfo  {journal} {Phys. Rev. D}\ }\textbf {\bibinfo {volume} {101}},\
  \bibinfo {pages} {076016} (\bibinfo {year} {2020})},\ \Eprint
  {http://arxiv.org/abs/2002.09757} {arXiv:2002.09757 [nucl-th]} \BibitemShut
  {NoStop}%
\bibitem [{\citenamefont {Li}\ \emph {et~al.}(2021)\citenamefont {Li},
  \citenamefont {Lappi},\ and\ \citenamefont {Zhao}}]{Li:2021zaw}%
  \BibitemOpen
  \bibfield  {author} {\bibinfo {author} {\bibfnamefont {M.}~\bibnamefont
  {Li}}, \bibinfo {author} {\bibfnamefont {T.}~\bibnamefont {Lappi}}, \ and\
  \bibinfo {author} {\bibfnamefont {X.}~\bibnamefont {Zhao}},\ }\href {\doibase
  10.1103/PhysRevD.104.056014} {\bibfield  {journal} {\bibinfo  {journal}
  {Phys. Rev. D}\ }\textbf {\bibinfo {volume} {104}},\ \bibinfo {pages}
  {056014} (\bibinfo {year} {2021})},\ \Eprint
  {http://arxiv.org/abs/2107.02225} {arXiv:2107.02225 [hep-ph]} \BibitemShut
  {NoStop}%
\bibitem [{\citenamefont {Li}\ \emph {et~al.}(2023)\citenamefont {Li},
  \citenamefont {Lappi}, \citenamefont {Zhao},\ and\ \citenamefont
  {Salgado}}]{Li:2023jeh}%
  \BibitemOpen
  \bibfield  {author} {\bibinfo {author} {\bibfnamefont {M.}~\bibnamefont
  {Li}}, \bibinfo {author} {\bibfnamefont {T.}~\bibnamefont {Lappi}}, \bibinfo
  {author} {\bibfnamefont {X.}~\bibnamefont {Zhao}}, \ and\ \bibinfo {author}
  {\bibfnamefont {C.~A.}\ \bibnamefont {Salgado}},\ }\href@noop {} {\
  (\bibinfo {year} {2023})},\ \Eprint {http://arxiv.org/abs/2305.12490}
  {arXiv:2305.12490 [hep-ph]} \BibitemShut {NoStop}%
\bibitem [{\citenamefont {Brodsky}\ \emph {et~al.}(1998)\citenamefont
  {Brodsky}, \citenamefont {Pauli},\ and\ \citenamefont
  {Pinsky}}]{Brodsky:1997de}%
  \BibitemOpen
  \bibfield  {author} {\bibinfo {author} {\bibfnamefont {S.~J.}\ \bibnamefont
  {Brodsky}}, \bibinfo {author} {\bibfnamefont {H.-C.}\ \bibnamefont {Pauli}},
  \ and\ \bibinfo {author} {\bibfnamefont {S.~S.}\ \bibnamefont {Pinsky}},\
  }\href {\doibase 10.1016/S0370-1573(97)00089-6} {\bibfield  {journal}
  {\bibinfo  {journal} {Phys. Rept.}\ }\textbf {\bibinfo {volume} {301}},\
  \bibinfo {pages} {299} (\bibinfo {year} {1998})},\ \Eprint
  {http://arxiv.org/abs/hep-ph/9705477} {arXiv:hep-ph/9705477} \BibitemShut
  {NoStop}%
\bibitem [{\citenamefont {McLerran}\ and\ \citenamefont
  {Venugopalan}(1994{\natexlab{a}})}]{McLerran:1993ka}%
  \BibitemOpen
  \bibfield  {author} {\bibinfo {author} {\bibfnamefont {L.~D.}\ \bibnamefont
  {McLerran}}\ and\ \bibinfo {author} {\bibfnamefont {R.}~\bibnamefont
  {Venugopalan}},\ }\href {\doibase 10.1103/PhysRevD.49.3352} {\bibfield
  {journal} {\bibinfo  {journal} {Phys. Rev. D}\ }\textbf {\bibinfo {volume}
  {49}},\ \bibinfo {pages} {3352} (\bibinfo {year} {1994}{\natexlab{a}})},\
  \Eprint {http://arxiv.org/abs/hep-ph/9311205} {arXiv:hep-ph/9311205}
  \BibitemShut {NoStop}%
\bibitem [{\citenamefont {McLerran}\ and\ \citenamefont
  {Venugopalan}(1994{\natexlab{b}})}]{McLerran:1993ni}%
  \BibitemOpen
  \bibfield  {author} {\bibinfo {author} {\bibfnamefont {L.~D.}\ \bibnamefont
  {McLerran}}\ and\ \bibinfo {author} {\bibfnamefont {R.}~\bibnamefont
  {Venugopalan}},\ }\href {\doibase 10.1103/PhysRevD.49.2233} {\bibfield
  {journal} {\bibinfo  {journal} {Phys. Rev. D}\ }\textbf {\bibinfo {volume}
  {49}},\ \bibinfo {pages} {2233} (\bibinfo {year} {1994}{\natexlab{b}})},\
  \Eprint {http://arxiv.org/abs/hep-ph/9309289} {arXiv:hep-ph/9309289}
  \BibitemShut {NoStop}%
\bibitem [{\citenamefont {Krasnitz}\ \emph {et~al.}(2001)\citenamefont
  {Krasnitz}, \citenamefont {Nara},\ and\ \citenamefont
  {Venugopalan}}]{Krasnitz:2001qu}%
  \BibitemOpen
  \bibfield  {author} {\bibinfo {author} {\bibfnamefont {A.}~\bibnamefont
  {Krasnitz}}, \bibinfo {author} {\bibfnamefont {Y.}~\bibnamefont {Nara}}, \
  and\ \bibinfo {author} {\bibfnamefont {R.}~\bibnamefont {Venugopalan}},\
  }\href {\doibase 10.1103/PhysRevLett.87.192302} {\bibfield  {journal}
  {\bibinfo  {journal} {Phys. Rev. Lett.}\ }\textbf {\bibinfo {volume} {87}},\
  \bibinfo {pages} {192302} (\bibinfo {year} {2001})},\ \Eprint
  {http://arxiv.org/abs/hep-ph/0108092} {arXiv:hep-ph/0108092} \BibitemShut
  {NoStop}%
\bibitem [{\citenamefont {Feynman}(1982)}]{Feynman:1981tf}%
  \BibitemOpen
  \bibfield  {author} {\bibinfo {author} {\bibfnamefont {R.~P.}\ \bibnamefont
  {Feynman}},\ }\href {\doibase 10.1007/BF02650179} {\bibfield  {journal}
  {\bibinfo  {journal} {Int. J. Theor. Phys.}\ }\textbf {\bibinfo {volume}
  {21}},\ \bibinfo {pages} {467} (\bibinfo {year} {1982})}\BibitemShut
  {NoStop}%
\bibitem [{\citenamefont {Zalka}(1998)}]{Zalka:1996st}%
  \BibitemOpen
  \bibfield  {author} {\bibinfo {author} {\bibfnamefont {C.}~\bibnamefont
  {Zalka}},\ }\href {\doibase 10.1098/rspa.1998.0162} {\bibfield  {journal}
  {\bibinfo  {journal} {Proc. Roy. Soc. Lond. A}\ }\textbf {\bibinfo {volume}
  {454}},\ \bibinfo {pages} {313} (\bibinfo {year} {1998})},\ \Eprint
  {http://arxiv.org/abs/quant-ph/9603026} {arXiv:quant-ph/9603026} \BibitemShut
  {NoStop}%
\bibitem [{\citenamefont {Wiesner}(1996)}]{Wiesner:1996xg}%
  \BibitemOpen
  \bibfield  {author} {\bibinfo {author} {\bibfnamefont {S.}~\bibnamefont
  {Wiesner}},\ }\href@noop {} {\  (\bibinfo {year} {1996})},\ \Eprint
  {http://arxiv.org/abs/quant-ph/9603028} {arXiv:quant-ph/9603028} \BibitemShut
  {NoStop}%
\bibitem [{\citenamefont {Nielsen}\ and\ \citenamefont
  {Chuang}(2010)}]{nielsen_chuang_2010}%
  \BibitemOpen
  \bibfield  {author} {\bibinfo {author} {\bibfnamefont {M.~A.}\ \bibnamefont
  {Nielsen}}\ and\ \bibinfo {author} {\bibfnamefont {I.~L.}\ \bibnamefont
  {Chuang}},\ }\href {\doibase 10.1017/CBO9780511976667} {\emph {\bibinfo
  {title} {Quantum Computation and Quantum Information: 10th Anniversary
  Edition}}}\ (\bibinfo  {publisher} {Cambridge University Press},\ \bibinfo
  {year} {2010})\BibitemShut {NoStop}%
\bibitem [{\citenamefont {Georgescu}\ \emph {et~al.}(2014)\citenamefont
  {Georgescu}, \citenamefont {Ashhab},\ and\ \citenamefont
  {Nori}}]{Georgescu:2013oza}%
  \BibitemOpen
  \bibfield  {author} {\bibinfo {author} {\bibfnamefont {I.~M.}\ \bibnamefont
  {Georgescu}}, \bibinfo {author} {\bibfnamefont {S.}~\bibnamefont {Ashhab}}, \
  and\ \bibinfo {author} {\bibfnamefont {F.}~\bibnamefont {Nori}},\ }\href
  {\doibase 10.1103/RevModPhys.86.153} {\bibfield  {journal} {\bibinfo
  {journal} {Rev. Mod. Phys.}\ }\textbf {\bibinfo {volume} {86}},\ \bibinfo
  {pages} {153} (\bibinfo {year} {2014})},\ \Eprint
  {http://arxiv.org/abs/1308.6253} {arXiv:1308.6253 [quant-ph]} \BibitemShut
  {NoStop}%
\bibitem [{\citenamefont {Barata}\ \emph {et~al.}(2021)\citenamefont {Barata},
  \citenamefont {Mueller}, \citenamefont {Tarasov},\ and\ \citenamefont
  {Venugopalan}}]{Barata:2020jtq}%
  \BibitemOpen
  \bibfield  {author} {\bibinfo {author} {\bibfnamefont {J.~a.}\ \bibnamefont
  {Barata}}, \bibinfo {author} {\bibfnamefont {N.}~\bibnamefont {Mueller}},
  \bibinfo {author} {\bibfnamefont {A.}~\bibnamefont {Tarasov}}, \ and\
  \bibinfo {author} {\bibfnamefont {R.}~\bibnamefont {Venugopalan}},\ }\href
  {\doibase 10.1103/PhysRevA.103.042410} {\bibfield  {journal} {\bibinfo
  {journal} {Phys. Rev. A}\ }\textbf {\bibinfo {volume} {103}},\ \bibinfo
  {pages} {042410} (\bibinfo {year} {2021})},\ \Eprint
  {http://arxiv.org/abs/2012.00020} {arXiv:2012.00020 [hep-th]} \BibitemShut
  {NoStop}%
\bibitem [{\citenamefont {Mueller}\ \emph {et~al.}(2020)\citenamefont
  {Mueller}, \citenamefont {Tarasov},\ and\ \citenamefont
  {Venugopalan}}]{Mueller:2019qqj}%
  \BibitemOpen
  \bibfield  {author} {\bibinfo {author} {\bibfnamefont {N.}~\bibnamefont
  {Mueller}}, \bibinfo {author} {\bibfnamefont {A.}~\bibnamefont {Tarasov}}, \
  and\ \bibinfo {author} {\bibfnamefont {R.}~\bibnamefont {Venugopalan}},\
  }\href {\doibase 10.1103/PhysRevD.102.016007} {\bibfield  {journal} {\bibinfo
   {journal} {Phys. Rev. D}\ }\textbf {\bibinfo {volume} {102}},\ \bibinfo
  {pages} {016007} (\bibinfo {year} {2020})},\ \Eprint
  {http://arxiv.org/abs/1908.07051} {arXiv:1908.07051 [hep-th]} \BibitemShut
  {NoStop}%
\bibitem [{\citenamefont {Yamawaki}(1998)}]{Yamawaki:1998cy}%
  \BibitemOpen
  \bibfield  {author} {\bibinfo {author} {\bibfnamefont {K.}~\bibnamefont
  {Yamawaki}},\ }in\ \href@noop {} {\emph {\bibinfo {booktitle} {{10th Summer
  School and Symposium on Nuclear Physics: QCD, Light cone Physics and Hadron
  Phenomenology (NuSS 97)}}}}\ (\bibinfo {year} {1998})\ pp.\ \bibinfo {pages}
  {116--199},\ \Eprint {http://arxiv.org/abs/hep-th/9802037}
  {arXiv:hep-th/9802037} \BibitemShut {NoStop}%
\bibitem [{\citenamefont {Deliyannis}\ \emph {et~al.}(2021)\citenamefont
  {Deliyannis}, \citenamefont {Freytsis}, \citenamefont {Nachman},\ and\
  \citenamefont {Bauer}}]{Deliyannis:2021che}%
  \BibitemOpen
  \bibfield  {author} {\bibinfo {author} {\bibfnamefont {P.}~\bibnamefont
  {Deliyannis}}, \bibinfo {author} {\bibfnamefont {M.}~\bibnamefont
  {Freytsis}}, \bibinfo {author} {\bibfnamefont {B.}~\bibnamefont {Nachman}}, \
  and\ \bibinfo {author} {\bibfnamefont {C.~W.}\ \bibnamefont {Bauer}},\
  }\href@noop {} {\  (\bibinfo {year} {2021})},\ \Eprint
  {http://arxiv.org/abs/2109.10918} {arXiv:2109.10918 [quant-ph]} \BibitemShut
  {NoStop}%
\bibitem [{\citenamefont {Kitaev}\ and\ \citenamefont
  {Webb}(2008)}]{KitaevGauss}%
  \BibitemOpen
  \bibfield  {author} {\bibinfo {author} {\bibfnamefont {A.}~\bibnamefont
  {Kitaev}}\ and\ \bibinfo {author} {\bibfnamefont {W.~A.}\ \bibnamefont
  {Webb}},\ }\href {\doibase 10.48550/ARXIV.0801.0342} {\enquote {\bibinfo
  {title} {Wavefunction preparation and resampling using a quantum computer},}\
  } (\bibinfo {year} {2008})\BibitemShut {NoStop}%
\bibitem [{\citenamefont {Lappi}(2008)}]{Lappi:2007ku}%
  \BibitemOpen
  \bibfield  {author} {\bibinfo {author} {\bibfnamefont {T.}~\bibnamefont
  {Lappi}},\ }\href {\doibase 10.1140/epjc/s10052-008-0588-4} {\bibfield
  {journal} {\bibinfo  {journal} {Eur. Phys. J. C}\ }\textbf {\bibinfo {volume}
  {55}},\ \bibinfo {pages} {285} (\bibinfo {year} {2008})},\ \Eprint
  {http://arxiv.org/abs/0711.3039} {arXiv:0711.3039 [hep-ph]} \BibitemShut
  {NoStop}%
\bibitem [{\citenamefont {Ipp}\ \emph {et~al.}(2020)\citenamefont {Ipp},
  \citenamefont {M\"uller},\ and\ \citenamefont {Schuh}}]{Ipp:2020mjc}%
  \BibitemOpen
  \bibfield  {author} {\bibinfo {author} {\bibfnamefont {A.}~\bibnamefont
  {Ipp}}, \bibinfo {author} {\bibfnamefont {D.~I.}\ \bibnamefont {M\"uller}}, \
  and\ \bibinfo {author} {\bibfnamefont {D.}~\bibnamefont {Schuh}},\ }\href
  {\doibase 10.1103/PhysRevD.102.074001} {\bibfield  {journal} {\bibinfo
  {journal} {Phys. Rev. D}\ }\textbf {\bibinfo {volume} {102}},\ \bibinfo
  {pages} {074001} (\bibinfo {year} {2020})},\ \Eprint
  {http://arxiv.org/abs/2001.10001} {arXiv:2001.10001 [hep-ph]} \BibitemShut
  {NoStop}%
\bibitem [{\citenamefont {Shende}\ \emph {et~al.}(2006)\citenamefont {Shende},
  \citenamefont {Bullock},\ and\ \citenamefont {Markov}}]{shende2006}%
  \BibitemOpen
  \bibfield  {author} {\bibinfo {author} {\bibfnamefont {V.}~\bibnamefont
  {Shende}}, \bibinfo {author} {\bibfnamefont {S.}~\bibnamefont {Bullock}}, \
  and\ \bibinfo {author} {\bibfnamefont {I.}~\bibnamefont {Markov}},\ }\href
  {\doibase 10.1109/TCAD.2005.855930} {\bibfield  {journal} {\bibinfo
  {journal} {IEEE Transactions on Computer-Aided Design of Integrated Circuits
  and Systems}\ }\textbf {\bibinfo {volume} {25}},\ \bibinfo {pages} {1000}
  (\bibinfo {year} {2006})}\BibitemShut {NoStop}%
\bibitem [{\citenamefont {Anis}\ \emph {et~al.}(2021)\citenamefont {Anis} \emph
  {et~al.}}]{Qiskit}%
  \BibitemOpen
  \bibfield  {author} {\bibinfo {author} {\bibfnamefont {M.~S.}\ \bibnamefont
  {Anis}} \emph {et~al.},\ }\href {\doibase {10.5281/zenodo.2573505}} {\enquote
  {\bibinfo {title} {{Qiskit: An Open-source Framework for Quantum
  Computing}},}\ } (\bibinfo {year} {{2021}})\BibitemShut {NoStop}%
\bibitem [{\citenamefont {Trotter}(1959)}]{Trotter:1959}%
  \BibitemOpen
  \bibfield  {author} {\bibinfo {author} {\bibfnamefont {H.~F.}\ \bibnamefont
  {Trotter}},\ }\href@noop {} {\bibfield  {journal} {\bibinfo  {journal}
  {Proceedings of the American Mathematical Society}\ }\textbf {\bibinfo
  {volume} {10}},\ \bibinfo {pages} {545} (\bibinfo {year} {1959})}\BibitemShut
  {NoStop}%
\bibitem [{\citenamefont {Suzuki}(1976)}]{Suzuki:1976}%
  \BibitemOpen
  \bibfield  {author} {\bibinfo {author} {\bibfnamefont {M.}~\bibnamefont
  {Suzuki}},\ }\href@noop {} {\bibfield  {journal} {\bibinfo  {journal}
  {Communications in Mathematical Physics}\ }\textbf {\bibinfo {volume} {51}},\
  \bibinfo {pages} {183} (\bibinfo {year} {1976})}\BibitemShut {NoStop}%
\bibitem [{\citenamefont {Hatano}\ and\ \citenamefont
  {Suzuki}(2005)}]{Hatano:2005gh}%
  \BibitemOpen
  \bibfield  {author} {\bibinfo {author} {\bibfnamefont {N.}~\bibnamefont
  {Hatano}}\ and\ \bibinfo {author} {\bibfnamefont {M.}~\bibnamefont
  {Suzuki}},\ }\href {\doibase 10.1007/11526216_2} {\bibfield  {journal}
  {\bibinfo  {journal} {Lect. Notes Phys.}\ }\textbf {\bibinfo {volume}
  {679}},\ \bibinfo {pages} {37} (\bibinfo {year} {2005})},\ \Eprint
  {http://arxiv.org/abs/math-ph/0506007} {arXiv:math-ph/0506007} \BibitemShut
  {NoStop}%
\bibitem [{\citenamefont {Subramanian}\ and\ \citenamefont
  {Hsieh}(2021)}]{Subramanian:2021}%
  \BibitemOpen
  \bibfield  {author} {\bibinfo {author} {\bibfnamefont {S.}~\bibnamefont
  {Subramanian}}\ and\ \bibinfo {author} {\bibfnamefont {M.-H.}\ \bibnamefont
  {Hsieh}},\ }\href {\doibase 10.1103/PhysRevA.104.022428} {\bibfield
  {journal} {\bibinfo  {journal} {Phys. Rev. A}\ }\textbf {\bibinfo {volume}
  {104}},\ \bibinfo {pages} {022428} (\bibinfo {year} {2021})}\BibitemShut
  {NoStop}%
\bibitem [{\citenamefont {Acharya}\ \emph {et~al.}(2020)\citenamefont
  {Acharya}, \citenamefont {Issa}, \citenamefont {Shende},\ and\ \citenamefont
  {Wagner}}]{Acharya:2020}%
  \BibitemOpen
  \bibfield  {author} {\bibinfo {author} {\bibfnamefont {J.}~\bibnamefont
  {Acharya}}, \bibinfo {author} {\bibfnamefont {I.}~\bibnamefont {Issa}},
  \bibinfo {author} {\bibfnamefont {N.~V.}\ \bibnamefont {Shende}}, \ and\
  \bibinfo {author} {\bibfnamefont {A.~B.}\ \bibnamefont {Wagner}},\ }\href
  {\doibase 10.1109/JSAIT.2020.3015235} {\bibfield  {journal} {\bibinfo
  {journal} {IEEE Journal on Selected Areas in Information Theory}\ }\textbf
  {\bibinfo {volume} {1}},\ \bibinfo {pages} {454} (\bibinfo {year}
  {2020})}\BibitemShut {NoStop}%
\bibitem [{\citenamefont {Li}\ and\ \citenamefont
  {Wu}(2019)}]{TongyangLi:2019}%
  \BibitemOpen
  \bibfield  {author} {\bibinfo {author} {\bibfnamefont {T.}~\bibnamefont
  {Li}}\ and\ \bibinfo {author} {\bibfnamefont {X.}~\bibnamefont {Wu}},\ }\href
  {\doibase 10.1109/TIT.2018.2883306} {\bibfield  {journal} {\bibinfo
  {journal} {IEEE Transactions on Information Theory}\ }\textbf {\bibinfo
  {volume} {65}},\ \bibinfo {pages} {2899} (\bibinfo {year}
  {2019})}\BibitemShut {NoStop}%
\bibitem [{\citenamefont {Dokshitzer}(1977)}]{Dokshitzer:1977sg}%
  \BibitemOpen
  \bibfield  {author} {\bibinfo {author} {\bibfnamefont {Y.~L.}\ \bibnamefont
  {Dokshitzer}},\ }\href@noop {} {\bibfield  {journal} {\bibinfo  {journal}
  {Sov. Phys. JETP}\ }\textbf {\bibinfo {volume} {46}},\ \bibinfo {pages} {641}
  (\bibinfo {year} {1977})},\ \bibinfo {note} {[Zh. Eksp. Teor.
  Fiz.73,1216(1977)]}\BibitemShut {NoStop}%
%%CITATION = SPHJA,46,641;%%
\bibitem [{\citenamefont {Altarelli}\ and\ \citenamefont
  {Parisi}(1977)}]{Altarelli:1977zs}%
  \BibitemOpen
  \bibfield  {author} {\bibinfo {author} {\bibfnamefont {G.}~\bibnamefont
  {Altarelli}}\ and\ \bibinfo {author} {\bibfnamefont {G.}~\bibnamefont
  {Parisi}},\ }\href {\doibase 10.1016/0550-3213(77)90384-4} {\bibfield
  {journal} {\bibinfo  {journal} {Nucl. Phys.}\ }\textbf {\bibinfo {volume}
  {B126}},\ \bibinfo {pages} {298} (\bibinfo {year} {1977})}\BibitemShut
  {NoStop}%
%%CITATION = NUPHA,B126,298;%%
\bibitem [{\citenamefont {Gribov}\ and\ \citenamefont
  {Lipatov}(1972)}]{Gribov:1972ri}%
  \BibitemOpen
  \bibfield  {author} {\bibinfo {author} {\bibfnamefont {V.~N.}\ \bibnamefont
  {Gribov}}\ and\ \bibinfo {author} {\bibfnamefont {L.~N.}\ \bibnamefont
  {Lipatov}},\ }\href@noop {} {\bibfield  {journal} {\bibinfo  {journal} {Sov.
  J. Nucl. Phys.}\ }\textbf {\bibinfo {volume} {15}},\ \bibinfo {pages} {438}
  (\bibinfo {year} {1972})},\ \bibinfo {note} {[Yad.
  Fiz.15,781(1972)]}\BibitemShut {NoStop}%
%%CITATION = SJNCA,15,438;%%
\bibitem [{\citenamefont {Ghiglieri}\ and\ \citenamefont
  {Weitz}(2022)}]{Ghiglieri:2022gyv}%
  \BibitemOpen
  \bibfield  {author} {\bibinfo {author} {\bibfnamefont {J.}~\bibnamefont
  {Ghiglieri}}\ and\ \bibinfo {author} {\bibfnamefont {E.}~\bibnamefont
  {Weitz}},\ }\href@noop {} {\  (\bibinfo {year} {2022})},\ \Eprint
  {http://arxiv.org/abs/2207.08842} {arXiv:2207.08842 [hep-ph]} \BibitemShut
  {NoStop}%
\bibitem [{\citenamefont {Caucal}\ and\ \citenamefont
  {Mehtar-Tani}(2021)}]{Caucal:2021lgf}%
  \BibitemOpen
  \bibfield  {author} {\bibinfo {author} {\bibfnamefont {P.}~\bibnamefont
  {Caucal}}\ and\ \bibinfo {author} {\bibfnamefont {Y.}~\bibnamefont
  {Mehtar-Tani}},\ }\href@noop {} {\  (\bibinfo {year} {2021})},\ \Eprint
  {http://arxiv.org/abs/2109.12041} {arXiv:2109.12041 [hep-ph]} \BibitemShut
  {NoStop}%
\bibitem [{\citenamefont {Liou}\ \emph {et~al.}(2013)\citenamefont {Liou},
  \citenamefont {Mueller},\ and\ \citenamefont {Wu}}]{Liou:2013qya}%
  \BibitemOpen
  \bibfield  {author} {\bibinfo {author} {\bibfnamefont {T.}~\bibnamefont
  {Liou}}, \bibinfo {author} {\bibfnamefont {A.~H.}\ \bibnamefont {Mueller}}, \
  and\ \bibinfo {author} {\bibfnamefont {B.}~\bibnamefont {Wu}},\ }\href
  {\doibase 10.1016/j.nuclphysa.2013.08.005} {\bibfield  {journal} {\bibinfo
  {journal} {Nucl. Phys. A}\ }\textbf {\bibinfo {volume} {916}},\ \bibinfo
  {pages} {102} (\bibinfo {year} {2013})},\ \Eprint
  {http://arxiv.org/abs/1304.7677} {arXiv:1304.7677 [hep-ph]} \BibitemShut
  {NoStop}%
\bibitem [{\citenamefont {Blaizot}\ and\ \citenamefont
  {Dominguez}(2019)}]{Blaizot:2019muz}%
  \BibitemOpen
  \bibfield  {author} {\bibinfo {author} {\bibfnamefont {J.-P.}\ \bibnamefont
  {Blaizot}}\ and\ \bibinfo {author} {\bibfnamefont {F.}~\bibnamefont
  {Dominguez}},\ }\href {\doibase 10.1103/PhysRevD.99.054005} {\bibfield
  {journal} {\bibinfo  {journal} {Phys. Rev. D}\ }\textbf {\bibinfo {volume}
  {99}},\ \bibinfo {pages} {054005} (\bibinfo {year} {2019})},\ \Eprint
  {http://arxiv.org/abs/1901.01448} {arXiv:1901.01448 [hep-ph]} \BibitemShut
  {NoStop}%
\bibitem [{\citenamefont {Barata}\ \emph {et~al.}(2023)\citenamefont {Barata},
  \citenamefont {Blaizot},\ and\ \citenamefont {Mehtar-Tani}}]{Barata:2023uoi}%
  \BibitemOpen
  \bibfield  {author} {\bibinfo {author} {\bibfnamefont {J.~a.}\ \bibnamefont
  {Barata}}, \bibinfo {author} {\bibfnamefont {J.-P.}\ \bibnamefont {Blaizot}},
  \ and\ \bibinfo {author} {\bibfnamefont {Y.}~\bibnamefont {Mehtar-Tani}},\
  }\href@noop {} {\  (\bibinfo {year} {2023})},\ \Eprint
  {http://arxiv.org/abs/2305.10476} {arXiv:2305.10476 [hep-ph]} \BibitemShut
  {NoStop}%
\bibitem [{\citenamefont {Arnold}\ \emph {et~al.}(2023)\citenamefont {Arnold},
  \citenamefont {Elgedawy},\ and\ \citenamefont {Iqbal}}]{Arnold:2023qwi}%
  \BibitemOpen
  \bibfield  {author} {\bibinfo {author} {\bibfnamefont {P.}~\bibnamefont
  {Arnold}}, \bibinfo {author} {\bibfnamefont {O.}~\bibnamefont {Elgedawy}}, \
  and\ \bibinfo {author} {\bibfnamefont {S.}~\bibnamefont {Iqbal}},\
  }\href@noop {} {\  (\bibinfo {year} {2023})},\ \Eprint
  {http://arxiv.org/abs/2302.10215} {arXiv:2302.10215 [hep-ph]} \BibitemShut
  {NoStop}%
\bibitem [{\citenamefont {Qian}\ \emph {et~al.}(2022)\citenamefont {Qian},
  \citenamefont {Basili}, \citenamefont {Pal}, \citenamefont {Luecke},\ and\
  \citenamefont {Vary}}]{Qian:2021jxp}%
  \BibitemOpen
  \bibfield  {author} {\bibinfo {author} {\bibfnamefont {W.}~\bibnamefont
  {Qian}}, \bibinfo {author} {\bibfnamefont {R.}~\bibnamefont {Basili}},
  \bibinfo {author} {\bibfnamefont {S.}~\bibnamefont {Pal}}, \bibinfo {author}
  {\bibfnamefont {G.}~\bibnamefont {Luecke}}, \ and\ \bibinfo {author}
  {\bibfnamefont {J.~P.}\ \bibnamefont {Vary}},\ }\href {\doibase
  10.1103/PhysRevResearch.4.043193} {\bibfield  {journal} {\bibinfo  {journal}
  {Phys. Rev. Res.}\ }\textbf {\bibinfo {volume} {4}},\ \bibinfo {pages}
  {043193} (\bibinfo {year} {2022})},\ \Eprint
  {http://arxiv.org/abs/2112.01927} {arXiv:2112.01927 [quant-ph]} \BibitemShut
  {NoStop}%
\bibitem [{\citenamefont {Romero}\ and\ \citenamefont
  {Santos-Su\'arez}(2023)}]{Romero:2023mob}%
  \BibitemOpen
  \bibfield  {author} {\bibinfo {author} {\bibfnamefont {S.~V.}\ \bibnamefont
  {Romero}}\ and\ \bibinfo {author} {\bibfnamefont {J.}~\bibnamefont
  {Santos-Su\'arez}},\ }\href@noop {} {\  (\bibinfo {year} {2023})},\ \Eprint
  {http://arxiv.org/abs/2301.00560} {arXiv:2301.00560 [quant-ph]} \BibitemShut
  {NoStop}%
\bibitem [{\citenamefont {Hadfield}(2021)}]{hadfield:2021}%
  \BibitemOpen
  \bibfield  {author} {\bibinfo {author} {\bibfnamefont {S.}~\bibnamefont
  {Hadfield}},\ }\href@noop {} {\bibfield  {journal} {\bibinfo  {journal} {ACM
  Transactions on Quantum Computing}\ }\textbf {\bibinfo {volume} {2}},\
  \bibinfo {pages} {1} (\bibinfo {year} {2021})}\BibitemShut {NoStop}%
\end{thebibliography}%

\end{document}